\documentclass[preprint2]{emulateapj}

\shorttitle{Spatial Clustering of RASS-AGNs}
\shortauthors{Krumpe et al.}

\begin{document}

\def\mpch {$h^{-1}$ Mpc} 
\def\kpch {$h^{-1}$ kpc} 
\def\kms {km s$^{-1}$} 
\def\lcdm {$\Lambda$CDM } 
\def\xir {$\xi(r)$}
\def\wprp {$w_p(r_p)$}
\def\xisp {$\xi(r_p,\pi)$}
\def\xis {$\xi(s)$}
\def\rr {$r_0$}
\def\etal {et al.}

\title{The Spatial Clustering of {\em ROSAT} All-Sky Survey AGNs\\ I. The cross-correlation function with SDSS Luminous Red Galaxies}

\author{Mirko Krumpe\altaffilmark{1}, Takamitsu Miyaji\altaffilmark{2,1}, and Alison L. Coil\altaffilmark{1}}

\altaffiltext{1}{University of California, San Diego, Center for Astrophysics and
                Space Sciences, 9500 Gilman Drive, La Jolla, CA 92093-0424, USA}
\altaffiltext{2}{IAUNAM-E (Instituto de Astronom\'ia de la Universidad Nacional
                 Aut\'onoma de M\'exico, Ensenada), P.O. Box 439027, San Diego, 
                 CA 92143-9027, USA}

\email{mkrumpe@ucsd.edu}

\begin{abstract}

We investigate the clustering properties of $\sim$1550 broad-line active galactic nuclei (AGNs) 
at $\langle$$z$$\rangle$=0.25 detected in the {\em ROSAT} All-Sky Survey (RASS) 
through their measured cross-correlation 
function with $\sim$46000 Luminous Red Galaxies (LRGs) in the 
Sloan Digital Sky Survey.  By measuring the cross-correlation of our AGN sample with a 
larger tracer set of LRGs, we both minimize shot noise errors due to the 
relatively small AGN sample size and avoid systematic errors due to 
the spatially-varying Galactic absorption that would affect direct 
measurements of the auto-correlation function (ACF) of the AGN sample.
The measured ACF correlation length
for the total RASS-AGN sample 
($<L_{\rm X}^{0.1-2.4\,{\rm keV}}>=1.5 \times 10^{44}$ erg\,s$^{-1}$) is 
$r_{\rm 0}=4.3^{+0.4}_{-0.5}$ $h^{-1}$ Mpc and the slope $\gamma=1.7^{+0.1}_{-0.1}$. 
Splitting the sample 
into low and high $L_{\rm X}$ samples at $L_{\rm X}^{0.5-10\,{\rm keV}} = 10^{44}$ erg\,s$^{-1}$,
we detect an X-ray luminosity-dependence of the clustering amplitude at the $\sim$2.5$\sigma$ level.
The low $L_{\rm X}$ sample has $r_{\rm 0}=3.3^{+0.6}_{-0.8}$ $h^{-1}$ Mpc 
($\gamma=1.7^{+0.4}_{-0.3}$), which is similar to the correlation length of blue 
star-forming galaxies at low redshift. The high $L_{\rm X}$ sample
has $r_{\rm 0}=5.4^{+0.7}_{-1.0}$ $h^{-1}$ Mpc ($\gamma=1.9^{+0.2}_{-0.2}$), which is consistent 
with the clustering of red galaxies. From the observed clustering amplitude, we infer that 
the typical dark matter halo (DMH) mass harboring RASS-AGNs with broad optical emission lines is 
log $(M_{\rm DMH}/(h^{-1}\,M_\odot)) =12.6^{+0.2}_{-0.3}$, 11.8$^{+0.6}_{-\infty}$, 
13.1$^{+0.2}_{-0.4}$ for the total, low $L_{\rm X}$, and high $L_{\rm X}$
RASS-AGN samples, respectively.  
 
\end{abstract}



\section{Introduction}

Galaxies and active galactic nuclei (AGNs) are not distributed randomly 
in the universe. The small primordial fluctuations in the matter density 
field present in 
the very early universe have progressively grown through gravitational 
collapse to create the complex network of clusters, groups, filaments, and 
voids seen in the distribution of structure today.  Galaxies and AGNs, as well as
groups and clusters of galaxies, are believed to populate the collapsed dark 
matter halos (DMHs).  The clustering of galaxies and AGNs therefore reflects the spatial
distribution of dark matter in the universe. This allows clustering 
measurements to be used to derive cosmological parameters 
(e.g., \citealt{peacock_cole_2001}; \citealt{abazajian_zheng_2005}). However, 
these measurements also allow us to study the complex physics which governs 
the creation and evolution of galaxies and AGNs, as well as the co-evolution 
of galaxies and AGNs. The co-evolution scenario is motivated by the observed 
correlation between the mass of the central super-massive black hole (SMBH) and the stellar 
velocity dispersion in the bulge (\citealt{gebhardt_bender_2000}; 
\citealt{ferrarese_merritt_2000}), lending strong evidence to an interaction or 
feedback mechanism between the SMBH and the host galaxy. The specific 
form of the feedback mechanism, as well as the details of the AGN triggering, 
accretion, and fueling mechanisms, remains unclear.
Different cosmological simulations address possible scenarios for the
 co-evolution of AGNs and their host galaxies (e.g., \citealt{kauffmann_haehnelt_2000}; 
\citealt{dimatteo_springel_2005}; \citealt{cattaneo_dekel_2006}).
Large volume, high-resolution simulations that include physical prescriptions for
 galaxy evolution and AGN feedback 
make predictions for the spatial clustering and large-scale environments of AGNs and 
galaxies (\citealt{springel_white_2005}; \citealt{colberg_dimatteo_2008}; 
\citealt{bonoli_marulli_2009}). 
Observed clustering measurements of AGNs can be used to test these theoretical models,
put constraints on the feedback mechanisms, identify the properties of the AGN 
host galaxies, and understand the accretion processes onto SMBHs and their 
fueling mechanism.   

X-ray surveys allow us to identify AGN activity without contamination from 
the emission of the host galaxy, i.e., therefore efficiently detecting even low luminosity AGNs.
In the current era of deep and wide-area X-ray 
surveys with extensive spectroscopic follow-up, 
measurements of the three-dimensional (3D) clustering of AGN
in various redshift ranges are emerging (\citealt{coil_georgakakis_2009}; 
\citealt{gilli_daddi_2005}; \citealt{yang_mushotzky_2006}). 
However, our knowledge of AGN clustering in the low redshift universe $(z \lesssim 0.4$) 
is still poor, except for optically selected type II AGNs (\citealt{wake_miller_2004}; 
\citealt{li_kauffmann_2006}). 
This is due to the lack of observable comoving volume and the low number density of 
low-$z$ AGNs. 
Exceptionally large survey areas with good sensitivity are needed to 
acquire a sufficiently large number of objects for clustering measurements. 

To date, the {\em ROSAT} All-Sky Survey (RASS; \citealt{voges_aschenbach_1999}) 
is the most sensitive survey to have mapped the entire sky in X-rays.
Surveys with modern higher-sensitivity X-ray observatories such as {\em XMM-Newton} 
and {\em Chandra} cover much smaller areas of the sky (area: $\sim$0.1-10 deg$^2$).
Therefore, the available comoving volume from these deeper data sets 
is not sufficient to accurately measure AGN clustering at low redshifts $(z \lesssim 0.4$). 
Serendipitous surveys, such as extended ChAMP (\citealt{covey_agueros_2008}) and 
2XMM (\citealt{watson_schroeder_2009}), cover larger areas ($\sim$33
deg$^2$ and $\sim$360 deg$^2$, respectively). However, the large 
variations in sensitivity between different observations and 
the non-contiguous sky coverage make serendipitous surveys
unsuitable for wide-area clustering measurements.

A few studies have attempted to measure the auto-correlation 
function (ACF) of RASS-based AGN samples. Two-dimensional (2D) angular correlation functions 
(\citealt{akylas_georgantopoulos_2000}) do not require redshift measurements for each
AGN, but the projection heavily dilutes the clustering signal. Furthermore, the 
deprojection of the angular clustering to the 3D correlation function 
is subject to uncertainties in the redshift distribution, which can be substantial. 
Direct measurements of the 3D RASS-AGN ACF with 
spectroscopic redshift measurements have been made by \cite{mullis_henry_2004} and 
\cite{grazian_negrello_2004} with a few hundred AGNs, respectively, providing 
clustering measurements that have large statistical uncertainties caused by 
the relatively small sample size.  

\cite{anderson_voges_2003, anderson_margon_2007} positionally cross-correlated 
RASS sources with spectroscopic data available from the Sloan Digital Sky Survey (SDSS). 
This dramatically increased the number of RASS-AGNs with 
spectroscopic redshift measurements, which we use here to 
provide significant improvements in the measurement of AGN clustering at low redshift.
Furthermore, the availability of spectroscopic redshifts for large samples of 
SDSS galaxies in the 
same volume allows us to use an alternative approach to infer the clustering of AGN 
using calculations of the AGN--galaxy cross-correlation function (CCF).
This approach uses much larger samples of AGN--galaxy pairs and hence significantly reduces 
the uncertainties in the spatial correlation function compared with direct measurements 
of the AGN ACF. Furthermore, the use of a CCF avoids the 
problem that we have to correct for the complex angular dependences of limiting 
sensitivity in the X-ray sample.

We therefore have initiated a program to investigate the clustering 
properties of low redshift ($z\sim 0.25$) RASS-AGNs through measurements of the 
CCF of these AGNs with SDSS Luminous Red Galaxies (LRGs). In this 
study, we chose LRGs as the corresponding galaxy sample because they have a 
significant overlap in redshift range with our X-ray sample 
(details are described later).

In this first paper of a series, we explain the data selection, as well as the 
calculation of the CCF and the inferred RASS-AGN ACF. We also investigate the X-ray 
luminosity dependence of the clustering properties and biases. In a follow-up paper 
(T. Miyaji et al., in preparation), we will focus on applying the halo occupation 
distribution (HOD) model to the calculated CCF between RASS-AGNs and LRGs. 

The paper is organized as follows. In Section~2, we describe the construction 
and properties of the LRG and X-ray AGN samples in details. All essential 
steps to measure the CCF, compute the ACF via the CCF, and estimate errors
are explained in Section~3. The results of the clustering measurements for the 
different X-ray AGN samples and their luminosity 
dependence are given in Section~4. We discuss these results in Section~5 in the context 
of other studies and conclude in Section~6. 
Throughout the paper, all distances, luminosities, and absolute magnitudes 
are measured in comoving coordinates and given in units of $h^{-1}$\,Mpc, 
where $h= H_{\rm 0}/100$\,km\,s$^{-1}$, unless otherwise stated. We use a 
cosmology of $\Omega_{\rm M} = 0.3$ and $\Omega_{\rm \Lambda} = 0.7$  (\citealt{spergel_verde_2003}). 
We use AB magnitudes throughout the paper. 
All uncertainties represent a 1$\sigma$ (68.3\%) confidence interval unless 
otherwise stated.


\section{Data}


\subsection{SDSS Luminous Red Galaxy Sample\label{LRG}}
The optical data analyzed in this study are drawn from the 
SDSS (\citealt{york_adelman_2000}).
We use data both from the main galaxy sample, which has a 
spectroscopic depth of 
$r=17.7$ (\citealt{strauss_weinberg_2002}), and the  
LRG sample, which has a spectroscopic
depth of $r=19.5$, significantly fainter than the main galaxy sample. 
The LRG sample was designed for studies of large-scale 
structure to higher redshift; it covers a larger volume than the main 
galaxy sample. 
Here we use the LRG sample as a large-scale structure tracer set to calculate 
the CCF with the RASS-AGN, as the LRG sample
covers a similar redshift range as the RASS-AGNs.

The target selection and the properties of the 
LRG sample are described in detail in \cite{eisenstein_annis_2001}. 
Two different selection criteria ('cut I' and 'cut II') were 
introduced in identifying LRGs as at $z \gtrsim 0.4$ the typical 
4000 \AA\ break in 
the spectral energy distribution (SED) of an early-type galaxy drops 
out of the $g$ band and falls into the $r$ band. \cite{eisenstein_annis_2001}
shows that up to $z\sim 0.38$ a volume-limited sample of LRGs with 
passively evolving luminosity and rest-frame colors is selected with a 
very high efficiency (95\% for cut I and 90\% for cut II). 
\cite{eisenstein_annis_2001} advised that the LRG selection algorithm 
should not used for objects $z < 0.15$.


\subsubsection{Extraction of the SDSS Luminous Red Galaxy Sample\label{extractionLRG}}
The LRG sample was obtained using the web-based SDSS Catalog Archive Server 
Jobs System\footnote{\tt http://casjobs.sdss.org/CasJobs/}. The appropriate 
objects are selected through the prime target flag 'galaxy\_red' 
(\citealt{eisenstein_annis_2001}). In SDSS DR2, the model magnitude code was 
changed (\citealt{abazajian_adelman-mccarthy_2004}) to improve the star--galaxy 
separation. This also caused a slight change in the LRG sample definition. We 
make use of the updated LRG sample selection.

We extract 115,577 LRGs in the DR7 data release
that have a spectroscopic redshift of $z>0.15$, a galaxy spectral-type
classification, and a redshift confidence level of $z_{\rm conf.} >0.95$. 
We discovered that 3\% of objects flagged 'galaxy\_red' (i.e., both cut selections) 
in DR7 do not fall within the LRG selection criteria (see \citealt{eisenstein_annis_2001}). 
This does not happen in earlier data releases prior to DR7. 

To calculate the RASS-AGN--LRG CCF, we wish to define an LRG sample that is 
both volume-limited and contains a high number density of LRGs to maximize
the number of possible AGN--LRG pairs.  
Table~1 of \cite{zehavi_eisenstein_2005} defines three different 
volume-limited LRG samples, corresponding to different luminosity ranges, 
that they use to measure the LRG ACF. 
The first volume-limited LRG sample, with $-23.2<M_g<-21.2$ and 
$0.16<z<0.36$, contains the highest number density 
of objects. Here we adopt the same LRG sample definition, as this sample
is the best suited to our scientific goals.
We select LRGs from DR7 that meet the criteria of this sample, and we
refer to the subsequent data here as 'LRG sample'. We further limited 
our survey area to DR4+ to match the X-ray sample 
(see Section~\ref{unification}). The properties of this sample are shown in 
Table~\ref{xagn_samples}.
     
The selected LRGs span a range in redshift. To construct a volume-limited sample, 
one must correct their time-evolving SED to account for the evolution 
of the stellar population and the redshifting.
As described in the Appendix of \cite{eisenstein_annis_2001}, we use 
the extinction corrected $r^{*}_{\rm petro}$ magnitude to construct the 
$k$-corrected and passively evolved rest-frame $g^{*}_{\rm petro}$ magnitudes 
(non-star-forming model). \cite{zehavi_eisenstein_2005} passively evolve 
their $M_g$ to $z=0.3$ instead of $z=0$ (\citealt{eisenstein_annis_2001}; 
we refer to this magnitude by using the notation $M_{g}^{z=0.3}$). The 
stellar population synthesis (SPS) code described in \cite{conroy_gunn_2008} 
is used to derive the purely passive evolution correction. LRGs are 
$\Delta M_g = -0.27$ brighter at $z=0.3$ compared to $z=0$. This value is 
similar to the one used in \cite{zehavi_eisenstein_2005} who state that 
the evolution is about 1 mag per unit redshift.

We extensively tested that our LRG sample meet the LRG 
selection criteria of both \cite{eisenstein_annis_2001} and 
\cite{zehavi_eisenstein_2005}. We only consider objects that are 
located in areas of the sky that have a spectroscopic completeness fraction 
in DR7 of $f_{\rm compl} > 0.8$.  The vast majority of the objects in our 
sample (99.9\%) belong to the LRG 'cut I' criteria.


\subsubsection{Accounting for the SDSS Fiber Collision\label{fibercollision}}
\cite{blanton_lin_2003} describe the algorithm used for positioning tiles in 
the plane of the sky and assigning spectroscopic fibers for observing 
objects in the SDSS. 
The resulting spectroscopic sample of SDSS is spatially biased in that 
fibers on the same tile cannot be placed closer than 55 arcsec. Spectroscopic 
redshifts are available for galaxy pairs closer than 55 arcsec only if the field is 
observed at least twice. As discussed in 
\cite{zehavi_blanton_2002,zehavi_eisenstein_2005}, not taking into account 
the effects of these fiber collisions would systematically underestimate 
measured correlation functions, even on large scales. 
\cite{blanton_schlegel_2005} provide publicly available collision-corrected 
SDSS catalogs suitable for robust large-scale structure studies. They assign a redshift
to a galaxy which has no spectroscopic redshift by giving it the redshift 
of that galaxy in a galaxy group that is positionally the closest and has a 
spectroscopic redshift. However, these catalogs are 
generated only for the main galaxy sample, not the LRG sample.

To correct our LRG sample for fiber collisions, 
we therefore use the following approach. We select from the SDSS 
catalog Archive Server 
all LRG objects that passed the pure photometric-based cut I and cut II 
selection criteria, with no spectroscopic restrictions applied. 
Photometrically selected LRGs that get a redshift assigned are only those 
which have a spectroscopic LRG within a separation of  $d <55\farcs0$.
This corresponds to 2\% (1004 objects) of the total LRG sample.
Our redshift assignment is a two stage process. First, if 
the photometrically selected LRG is included in the primary spectroscopic
LRG sample but its redshift was initially rejected due to 
a low confidence level of $z_{\rm conf.} \le 0.95$, the spectroscopic 
redshift is now accepted independent of its confidence level to correct 
for fiber collisions. This specific procedure assigns a redshift only 
to nine objects. 

As a next step, we make use of the substantially improved 
DR7 photometric redshift code (\citealt{abazajian_adelman-mccarthy_2008}) 
compared to earlier SDSS data releases. 
If the difference between the spectroscopic redshift ($z_{{\rm spec,}j}$) of 
an LRG and the photometric redshift ($z_{{\rm photo,}i}$) of a neighboring 
(within a 55 arcsecond radius) LRG is 
\begin{equation}\label{eq:redshift}
\mid z_{{\rm spec,}j} - z_{{\rm photo,}i} \mid \le \delta z_{{\rm photo,}i,1\sigma},  
\end{equation}
then the photometric LRG is assigned the same redshift as the spectroscopic 
LRG. Otherwise the photometric redshift is taken to be the correct redshift 
for the photometric 
LRG. More than half of all photometric LRG redshifts are assigned the 
spectroscopic redshift of a neighboring LRG, while the remaining objects 
are assigned their photometric redshift. 

The above procedure for redshift assignment is used because simply adopting
the photometric redshift for all objects that do not have a measured 
spectroscopic redshift would smear out the clustering signal in redshift space 
and therefore dilute the amplitude of the correlation function. 
\cite{blanton_lin_2003} verified that for areas of the sky with overlapping
spectroscopic tiles, 60\% of the galaxies that had a neighboring galaxy within
55 arcsec were within 10 Mpc of the neighbor galaxy.
\cite{zehavi_eisenstein_2005} demonstrated that LRGs 
are highly clustered, so using this procedure for LRGs will result in a 
higher success rate.
The average photometric redshift error ($\delta z/(1+z)$) is 2\%, 
and within our sample less than 1\% (437 objects) of the LRGs are assigned
a photometric redshift.

\subsection{The X-ray Sample}
The RASS (\citealt{voges_aschenbach_1999}) 
is currently the most sensitive all-sky survey in the X-ray regime, with a
 typical flux limit of 
$f_{\rm X} \sim 10^{-13}\,\rm{erg}\,\rm{cm}^{-2}\,\rm{s}^{-1}$ (0.1-2.4 keV).
The area of the sky covered is 99.7\%, all of which has at least 50 s of 
exposure with {\em ROSAT} Position Sensitive Proportional Counters (PSPC). 

Based on the SDSS data release 5 (DR5), \cite{anderson_margon_2007}
classify 6224 AGNs with broad permitted emission lines in excess 
of 1000 km\,s$^{-1}$ FWHM and 515 narrow permitted emission line AGNs 
matching RASS sources within 1 arcmin 
(note that 86\% of all matches fall within 30$\farcs0$). Consequently, 
broad-line AGNs account for 92\% of all RASS/SDSS classifications. In this study, 
we use the broad emission line RASS-AGNs only.
The RASS sources
are taken from both the RASS Bright source catalog (\citealt{voges_aschenbach_1999}) 
and the RASS Faint source catalog (\citealt{voges_aschenbach_2000}) and have    
a maximum likelihood ML $ \ge 7$.\footnote{\cite{anderson_voges_2003, anderson_margon_2007} 
say that sources with ML $ \ge 10$ have been included. However, investigating 
the sample shows that ML $ \ge 7$ has been used.} The data cover an area 
of 5740 deg$^2$. The derived luminosities in \cite{anderson_margon_2007}
are based on a flat $\Lambda$CDM cosmology with 
($\Omega_{\rm M}$, $\Omega_{\Lambda}$, $H_0$)=(0.3, 0.7, 70\,km\,s$^{-1}$\,Mpc$^{-1}$).
\begin{figure}
  \centering
 \resizebox{\hsize}{!}{ 
  \includegraphics[bbllx=85,bblly=369,bburx=540,bbury=700]{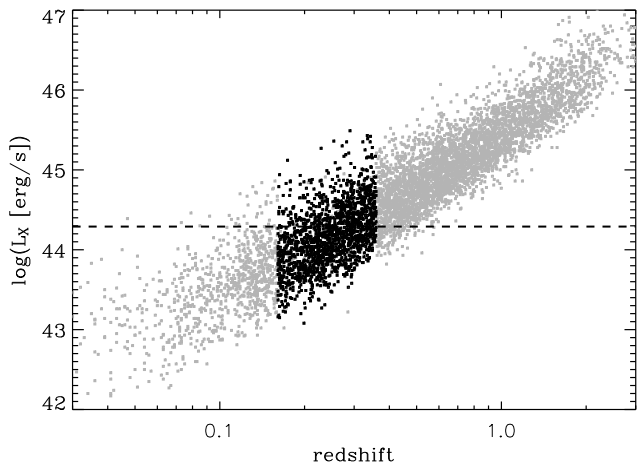}} 
      \caption{0.1-2.4 keV X-ray luminosity vs. redshift for 
        the broad emission line AGN sample in SDSS DR5 determined by
               \cite{anderson_margon_2007}. Black symbols show objects 
               in the redshift range 
               $0.16 < z < 0.36$, which define the AGN sample used here. 
               The dashed line corresponds to the common 
               dividing line between low luminous AGNs and high luminous QSOs
               ($L_{\rm X}^{0.5-10\,{\rm keV}} = 10^{44}$\,erg\,s$^{-1}$) defining
               the low and high luminosity RASS-AGN samples used here.}
         \label{LX_z}
\end{figure}

The RASS/SDSS target selection consists of two selection algorithm: 
the main SDSS algorithm for galaxies and AGNs ($15.0 < m < 19.0$) and the 
specific assignment of SDSS fibers for RASS identifications ($15.0 < m < 20.5$).
\cite{anderson_voges_2003} explain in detail their selection of optical 
counterpart candidates.

To understand possible bias effects, which may influence our CCF measurements, 
we summarize the selection method briefly. The algorithm considers 
those SDSS optical objects within 1 arcmin of the X-ray source position. 
Optically detected objects qualify for SDSS spectroscopy if their $g$, $r$, 
or $i$ magnitude is $15.0 < m < 20.5$. Since not every optically detected 
object that fulfills the criteria can be observed spectroscopically, different 
priority levels are introduced in target selection. 
The highest priority is given to objects with a triple 
positional coincidence of RASS X-ray source, SDSS optical object, and FIRST 
(\citealt{becker_white_1995}) radio source. The next priority level
consists of SDSS objects in the RASS error circles with unusual optical colors 
indicating AGNs (e.g., UV-excess with $u-g<0.6$). The third priority 
includes less likely, but plausible, counterparts such as bright stars and 
galaxies. The last group consists of any object in the magnitude limit 
that falls within the RASS error circle. One RASS/SDSS spectroscopic fiber is 
assigned to the object with the highest priority level and a spatial 
offset less than 27.5 arcsec from the X-ray source. If multiple objects 
populate the same priority level, one is chosen at random for SDSS spectroscopy. 

Before consideration is given to potential RASS targets, an SDSS fiber is 
first assigned to objects that belong to the main SDSS sample, which includes 
both galaxies and optically identified AGNs. About 80\% of all RASS/SDSS 
identifications are independently targeted for spectroscopy by the SDSS main 
target selection algorithm. 
However, objects in the main sample can be observed at the cost of excluding 
high-priority RASS targets. In these cases, the RASS/SDSS identification is missing.
For these reasons, the RASS/SDSS AGN sample is only reasonably complete 
($\gtrsim 90$\%) for AGNs with $15.0 < m < 19.0$.

Objects identified by RASS/SDSS priority levels and not because of their 
belonging to the main SDSS sample tend to be optically fainter, on 
average, than the main AGN sample. In the $u-g$ versus $g-r$ color--color 
space, objects from both selection algorithms lie in the same region. 
Studying their redshift distributions shows that objects identified by the 
RASS/SDSS priority levels are, on average, at higher redshifts than the 
identifications relying on the main SDSS sample. This also
explains their lower observed count rates X-ray fluxes. The X-ray hardness 
ratios of both populations are very similar. Being fainter in X-rays and 
in the optical wavelengths leads to the same X-ray to optical flux ratio 
for the RASS/SDSS priority levels AGNs and the main SDSS sample AGNs. 
Therefore, except that RASS/SDSS priority levels AGNs are found on average at 
higher redshifts, no intrinsic differences between the RASS/SDSS AGNs from 
both selection methods are detected. 

To create a well-defined X-ray AGN sample to use for clustering measurements, 
we focus solely on broad emission line AGNs. {\em ROSAT}'s sensitivity 
to the soft energy band that is limited to $<$2.4 keV selects against 
X-ray absorbed AGNs (type II), 
which are usually optically classified as narrow lines objects. 
Therefore, the sample of broad emission line RASS-AGN is much 
more complete than the narrow emission line RASS-AGN sample. 
Furthermore, the X-ray/optical counterpart identification is more reliable 
for the broad emission line AGNs than for the narrow emission line AGNs, which have 
much higher surface density (see \citealt{anderson_voges_2003, anderson_margon_2007}
for details).

In addition to studying the clustering of the RASS-AGN sample, here we also 
test for possible differences in the clustering of low versus high X-ray
luminosity RASS-AGN  (Figure~\ref{LX_z}). The commonly used X-ray
dividing line between Seyfert AGNs and high luminous QSOs is
\mbox{$L_{\rm X}^{0.5-10 \rm keV}= 10^{44}$ erg\,s$^{-1}$} in the
0.5-10 keV band (intrinsic luminosity;
\citealt{mainieri_bergeron_2002}). Although this dividing line is
somewhat arbitrary, it is widely employed in literature, and we use
it here.
\cite{anderson_margon_2007} provide a table which list the
galactic absorption-corrected 0.1-2.4 keV luminosities
assuming a photon index of $\Gamma=2.5$.  With this index, the
AGN/QSO dividing line corresponds to a 0.1-2.4 keV luminosity of
\mbox{log $(L_{\rm X}/(\rm{erg}\,\rm{s}^{-1}))=44.298$} (Figure~\ref{LX_z}).
This index agrees well with 
that found by \cite{piconcelli_jimenez_2005} 
in the {\em XMM-Newton} spectra of PG quasars in the 0.5-2 keV band, where
$\bar{\Gamma}=2.73^{+0.12}_{-0.11}$.  
In the same paper, it is shown that at higher energies (2-12 keV) the mean photon 
index for PG quasars gets harder and is $\bar{\Gamma}=1.89\pm 0.11$. Here 
we use $\Gamma=2.5$ to be consistent with \cite{anderson_margon_2007}.

The properties of our total RASS-AGN sample, low $L_{\rm X}$ RASS-AGN sample, 
and the high $L_{\rm X}$ RASS-AGN sample are shown in Table~\ref{xagn_samples}.
In order to measure the CCF of the RASS X-ray AGNs with LRGs,
the X-ray samples must cover the same area of sky and redshift range
as the LRG sample used here. 
In all samples, no X-ray detected AGN is also classified as an LRG. However, we cannot 
exclude the possibility that RASS-AGNs are hosted by LRGs and outshine their host galaxies. 
In contrast to 
the high $L_{\rm X}$ RASS-AGN sample, the total RASS-AGN sample and the low 
$L_{\rm X}$ RASS-AGN sample are not volume-limited. We calculate the comoving number density 
in Table~\ref{xagn_samples} of these two samples in the following way. 
For a specific R.A. and decl. (contained in the DR4+ geometry), we determine the 
galactic absorption value $N_{\rm H}$ and the RASS exposure time. The RASS faint 
source catalog contains sources with at least six source counts. The latter leads to a 
limiting observable count rate for a given R.A. and decl. Using 
{\tt Xspec}, we can compute the Galactic absorption-corrected flux limit versus survey area 
for the RASS-AGN based on count rates, $N_{\rm H}$ values, and $\Gamma=2.5$. We then compute  
the comoving volume ($z_{\rm low}=0.16$) available to each object ($V_{\rm a}$) for being 
included in 
the sample following \cite{avni_bahcall_1980}, using the Galactic absorption-corrected object fluxes from 
\cite{anderson_margon_2007}. From this we calculate the comoving number density as 
$n_{\rm AGN} = \sum_{i} 1/V_{{\rm a},i}$, where $i$ sums over each object.

\begin{figure*}[t]
  \centering
 \resizebox{\hsize}{!}{ 
  \includegraphics[bbllx=85,bblly=369,bburx=540,bbury=700]{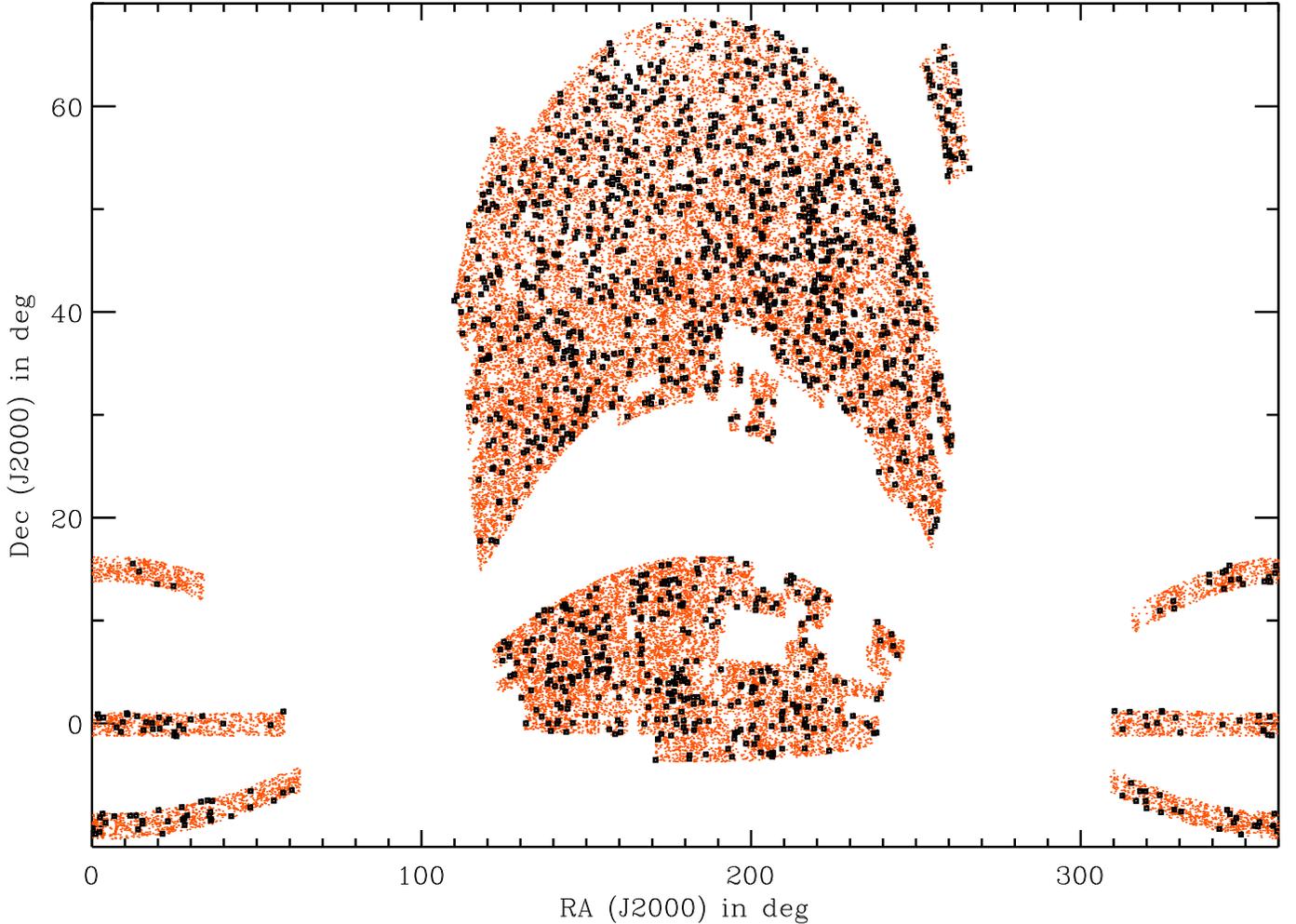}} 
      \caption{Spatial distribution of the total RASS-AGN sample (black squares) 
               and SDSS LRG sample used here (dots, our LRG sample) restricted to the 
               SDSS DR4+ geometry. The total covered area is 5468 deg$^2$.}
         \label{LRG_XAGN}
\end{figure*}

\begin{deluxetable*}{ccccccc}
\tabletypesize{\normalsize}
\tablecaption{Properties of the LRG and RASS-AGN Samples.\label{xagn_samples}}
\tablewidth{0pt}
\tablehead{
\colhead{Sample} & \colhead{}  & \colhead{$M_{g}^{z=0.3}$ Range} & \colhead{} & \colhead{$<n_{\rm LRG}>$} & \colhead{} & \colhead{$<M_{g}^{z=0.3}>$}\\
\colhead{Name} & \colhead{$z$-range}  & \colhead{(mag)} & \colhead{Number} & \colhead{($h^{3}$ Mpc$^{-3}$)} & \colhead{$\langle$$z$$\rangle$} & \colhead{(mag)}}
\startdata

LRG sample            & $0.16 <z<0.36$ & $-23.2 < M_{g}^{z=0.3} < -21.2$  & 45899 & $9.6 \times 10^{-5}$ & 0.28& -21.71 \\\hline
                      &                &                                &       &                      &     & \\    
                      &                & $L_{\rm X}^{0.1-2.4\,{\rm keV}}$           &       & $<n_{\rm AGN}>$       &     & $<L_{\rm X}^{0.1-2.4\,{\rm keV}}>$\\
                      & $z$-range      & Range (erg s$^{-1}$)           & Number& ($h^{3}$ Mpc$^{-3}$) &$\langle$$z$$\rangle$& (erg s$^{-1}$)\\\hline
Total RASS-AGN sample & $0.16 <z<0.36$ & --                             &  1552 & $6.0 \times 10^{-5}$ & 0.25 & $1.49\times 10^{44}$      \\
Low $L_{\rm X}$ RASS-AGN sample & $0.16 <z<0.36$ &$L_{\rm X} \le 1.95\times 10^{44}$& 990  & $5.8 \times 10^{-5}$ & 0.24 & $8.81\times 10^{43}$ \\
High $L_{\rm X}$ RASS-AGN sample & $0.16 <z<0.36$ &$L_{\rm X} > 1.95\times 10^{44}$  & 562 & $1.2 \times 10^{-6}$ & 0.28 & $3.78\times 10^{44}$
\enddata
\end{deluxetable*}


\subsection{Defining a Common Survey Geometry of the RASS-AGN and LRG Samples\label{unification}}
The RASS/SDSS \cite{anderson_margon_2007} sample is based on the SDSS DR5, while 
the LRG sample is drawn from SDSS DR7. We make use of DR7 for the LRG sample as 
it contains the latest 
available and furthest advanced version of the SDSS products.
Numerous correction have been applied in comparison to earlier data releases (see 
\citealt{abazajian_adelman-mccarthy_2008}; e.g., updated photo-$z$, 
repeated observations for few regions with poor seeing in previous data releases, filling 
holes in DR6 region, correction of instability in the spectroscopic flat-fields).
We then limit our LRG sample to the region covered by the \cite{anderson_margon_2007} AGN sample
for the CCF calculation.

The SDSS survey geometry and completeness are expressed in terms of 
spherical polygons (\citealt{hamilton_tegmark_2004}). Publicly available 
geometry and completeness files are not available for DR5, which would have 
the largest common survey area between the LRG and the 
RASS-AGN samples. Therefore, we use the latest version available prior to DR5: 
the DR4+ geometry file,\footnote{\tt http://sdss.physics.nyu.edu/lss/dr4plus}, 
which is a subset 
(\citealt{park_choi_2007}) of the SDSS DR5 (\citealt{adelman-mccarthy_agueeros_2007}) 
and covers 5540 deg$^2$ (DR5: 5740 deg$^2$).

The final, fiber collision-corrected LRG sample used here is based on DR7 but 
reconfigured to include only the DR4+ survey area that has DR7 completeness ratios 
of $f_{\rm compl} > 0.8$. The corresponding area of this sample is 5468 deg$^2$. 
This reconfiguration of the total \cite{anderson_margon_2007} sample from DR5 to DR4+ 
eliminates 287 broad emission line objects, leaving 5937 AGNs. Applying the 
redshift range selection of the LRG sample results in the number of objects 
given in Table~\ref{xagn_samples} for each AGN sample.
Figure~\ref{LRG_XAGN} shows the sky coverage of our final RASS-AGN sample and the 
LRG sample reconfigured to a 
common DR4+ geometry which is used for the calculation of the CCF.


\section{Measuring the Cross-correlation Function}

A commonly used technique for measuring the spatial clustering of a class of objects is 
the two-point 
 correlation function $\xi(r)$ (\citealt{peebles_1980}), which measures the excess 
probability $dP$ above a Poisson distribution of finding an object in a volume element $dV$
at a distance $r$ from another randomly chosen object:
\begin{equation}
\label{two-point_corr_f}
 dP =n[1+\xi(r)]dV,
\end{equation}
where $n$ is the mean number density of objects.
The ACF measures the excess probability of finding two objects from the same 
sample in a given volume element, 
while the CCF measures the excess probability finding an object 
from one sample at a distance $r$ from another object drawn 
from a different sample. The two-point correlation function, $\xi(r)$, is equal to 0 
for randomly distributed
objects, and $\xi(r)> 0$ if objects are more strongly clustered than a randomly distributed
sample.

In practice, the correlation function is obtained by counting pairs of objects with a given 
separation and comparing to the number of pairs in a random sample for the same 
separation. Different correlation estimators are described in the literature. 
\cite{davis_peebles_1983} give a simple estimator with the form
\begin{equation}
\label{DD_DR}
 \xi(r)= \frac{DD(r)}{DR(r)} -1,
\end{equation}
where $DD(r)$ is the sum of the data--data pairs at the separation $r$ and $DR(r)$ 
is the number data--random pairs; both pair counts have been normalized. 
\cite{landy_szalay_1993} suggest a more advanced estimator:
\begin{equation}
\label{DD_DR_RR}
 \xi=\frac{1}{RR}\left[DD-2DR+RR\right],
\end{equation} 
where $RR$ is the normalized number of random--random pairs; this estimator 
yields errors similar to what is expected for Poisson errors only. 

Because we measure line-of-sight distances from redshifts, the measurement of $\xi$ is 
subject to the redshift-space distortions due to peculiar velocities. To separate
the effects of redshift distortions, the spatial correlation function is measured as a 
function of two components of the separation vector between two objects, i.e., one 
perpendicular to ($r_p$) and the other along ($\pi$) the line of sight. Therefore, 
$\xi(r_P,\pi )$ is extracted by counting pairs 
on a 2D grid of separations $r_p$ and $\pi$. The real-space correlation function 
$\xi(r)$ can be recovered by integrating along the $\pi$ direction and computing the projected 
correlation function by \cite{davis_peebles_1983}
\begin{eqnarray}
\label{pi_integration}
w_p(r_p) &=& 2\int_0^{\infty}\!d\pi\,\xi(r_p,\pi) \nonumber\\
         &=& 2\int_0^{\infty}\!dy\,\xi \left[(r_p^2+y^2)^{(1/2)}\right] \nonumber\\
         &=& 2\int_{r_p}^{\infty}r\,\!dr\,\xi(r)\left( r^2-r_p^2 \right) ^{-1/2}
\end{eqnarray}
which is independent of redshift-space distortions. The variable $y$ represents the 
real-space separation along the line of sight. 
For a power law correlation function   
\begin{eqnarray}
 \xi(r)=\left(\frac{r}{r_{\rm 0}}\right)^{-\gamma},
\end{eqnarray}
$r_{\rm 0}$ and $\gamma$ are readily extracted from the projected correlation 
function using the analytical solution
\begin{eqnarray}
 w_p(r_p) &=& r_p\left(\frac{r_{\rm 0}}{r_p}\right)^{\gamma}\,\frac{\Gamma(1/2)\Gamma((\gamma-1)/2)}{\Gamma(\gamma/2)},
\end{eqnarray}
where $\Gamma(x)$ is the Gamma function.

The aim of this paper is to more accurately measure the clustering properties 
of low-$z$ AGNs than has been measured previously using ACFs
 (e.g., \citealt{mullis_henry_2004}; \citealt{grazian_negrello_2004}); we 
accomplish this by measuring the CCF of the AGNs with higher-density LRGs in the
same volume. Assuming 
a linear bias, we follow \cite{coil_georgakakis_2009} and infer the ACF of the 
AGN sample using 
\begin{eqnarray}
\label{acf_rassagn}
 w_p({\rm AGN}|{\rm AGN}) = \frac{\left[w_p({\rm AGN}|{\rm LRG})\right]^2}{w_p({\rm LRG}|{\rm LRG})}\,,
\end{eqnarray}
where $w_p({\rm AGN}|{\rm AGN})$ and $w_p({\rm LRG}|{\rm LRG})$ are the ACFs of the 
RASS-AGNs and the LRGs, respectively, and $w_p({\rm AGN}|{\rm LRG})$ is the 
CCF of the RASS-AGNs with the LRGs. The LRG ACF 
is studied extensively in \cite{zehavi_eisenstein_2005}, where the estimator in 
Equation~(\ref{DD_DR_RR}) is used; the results are given here in Table~\ref{table:lrg}. 
The RASS-AGN--LRG CCF is computed here using the estimator given in Equation~(\ref{DD_DR})
\begin{eqnarray}
 \xi_{\rm AGN-LRG} = \frac{D_{\rm AGN}\,D_{\rm LRG}}{D_{\rm AGN}\,R_{\rm LRG}}-1.
\end{eqnarray}
We use this estimator as it requires a 
random catalog for only the LRG sample ($R_{\rm LRG}$), which is homogenous, volume-limited,  
and has a well-understood selection function.  
The estimator given in Equation~(\ref{DD_DR_RR}) would require a random catalog for the RASS-AGN 
sample, which would be subject to possible systematic biases 
due to difficulties in accurately modeling the position-dependent sensitivity limit.
Especially, the changing Galactic absorption over the sky causes variations in the flux limit, 
which would require spectrum-dependent corrections.


\subsection{Construction of the Random LRG Sample\label{randomLRG}}
The generation of random samples is crucial for a proper measurements of the correlation function.
The objective is to construct a sample of randomly distributed sources that have the 
same observational survey biases as the real sample. Use of the estimator given 
in Equation~(\ref{DD_DR}) requires that the construction of a random sample for only the LRG
population is needed.

The LRG sample used here has been corrected for fiber collisions (Section~\ref{fibercollision});
therefore, we do not have to consider this bias in the construction of the random LRG sample.
The SDSS survey geometry and completeness ratio for a given field are given by 
the SDSS geometry files. For a set of random R.A. and decl. values, they allow us to determine 
if an object is covered by the SDSS DR4+ geometry and the spectroscopic completeness ratio at that
location.
Only objects covered in DR4+ with a DR7 spectroscopic completeness ratio $f_{\rm compl} > 0.8$ are 
accepted for the random sample required here. Additionally, the DR7 spectroscopic completeness 
is used as the probability 
that an object is kept for the random sample. If the completeness ratio is 0.9, the object has 
a 90\% 
chance of being included in the final random LRG sample. This procedure takes into account 
the fact that survey regions 
with a high spectroscopic completeness ratio have, on average, a higher object density than 
less complete areas. 

\begin{figure}
  \centering
 \resizebox{\hsize}{!}{ 
  \includegraphics[bbllx=76,bblly=369,bburx=540,bbury=700]{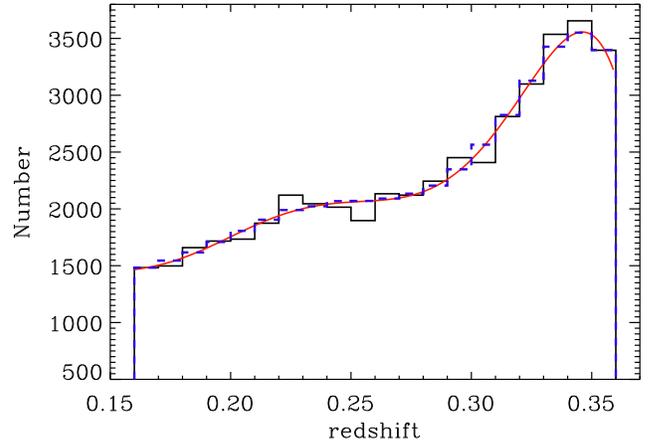}} 
      \caption{Redshift histogram of the LRG data sample (solid black lines), the smoothed profile 
        (red), and the random LRG sample (blue-dashed, renormalized to the total number of LRGs).}
              \label{z_histogram}
\end{figure}

The corresponding redshift for a random object is assigned based on the smoothed 
redshift distribution of the LRG data sample. The smoothing has been made by applying a 
least-squares (\citealt{savitzky_golay_1964}) low-pass filter to the 
observed LRG redshift distribution. We compare the redshift distribution of the LRG data sample, its smoothed profile, 
and the redshift distribution of the random LRG sample in Figure~\ref{z_histogram}. 
The shape of the LRG redshift distribution is caused by the superposition of two 
selection criteria for the LRGs. At low redshifts, most of the LRGs are selected in the SDSS
main galaxy sample. This flux-limited selection reaches the maximum at $z\sim 0.22$.
The 'cut I' LRG selection provides objects already from the lowest redshifts with a fast 
increasing number of objects for higher redshifts. This selection reaches its flux limit 
at $z\sim 0.35$. The 'cut II' selection plays an important role only at $z\gtrsim 0.42$. 
The selection dependence in different redshifts is illustrated in detail in Figure~12--14 of 
\cite{eisenstein_annis_2001}.

The random catalog contains 100 times as many objects as the LRG sample. This value is 
chosen to have an adequate number of pairs in the $D_{\rm AGN}\,R_{\rm LRG}$ sample at the 
smallest scales measured here. 

\subsection{Errors and Covariance Matrices}

The calculation of realistic error bars on measurements of the correlation function has been a 
subject of debate since the earliest measurements. Different methods 
are summarized in \cite{norberg_baugh_2009}. 
Adjacent bins in $w_p(r_p)$ are correlated, as are their errors.
The construction of a 
covariance matrix $M_{ij}$, which reflects the degree to which bin $i$ is correlated with 
bin $j$, is needed
to obtain meaningful power law fits to $w_p(r_p)$. 

We estimate the statistical errors of our correlation measurements using the jackknife 
method. We divide the survey area into $N_{\rm T}=100$ sections, each of which is $\sim$55 deg$^2$. 
We calculate $w_p(r_p)$ $N_{\rm T}$ times, where each jackknife sample excludes one section.
These $N_{\rm T}$ jackknife-resampled RASS-AGN ACFs are used to derive the 
covariance matrix $M_{ij}$ by  
\begin{eqnarray}
\label{jackknife}
 M_{ij} = \frac{N_{\rm T} -1}{N_{\rm T}} \left[\sum_{k=1}^{N_{\rm T}} \bigg(w_k(r_{p,i})-<w(r_{p,i})>\bigg)\right.\nonumber\\
          \times \bigg(w_k(r_{p,j})-<w(r_{p,j})>\bigg)\bigg] \, 
\end{eqnarray}
where $w_k(r_{p,i})$ and $w_k(r_{p,j})$ are from the $k$-th jackknife samples of the RASS-AGN ACF and 
$<w(r_{p,i})>$, $<w(r_{p,j})>$ are the averages over all of the jackknife samples. 
The 1$\sigma$ error of each bin is the square root of the diagonal component of this 
matrix ($\sigma_{i}=\sqrt{M_{ii}}$).
To calculate the covariance matrix of the RASS-AGN ACF, which is determined using 
Equation~(\ref{acf_rassagn}), we compute the RASS-AGN ACF for each of the $N_{\rm T}$ 
jackknife samples from the corresponding RASS-AGN--LRG CCF and LRG ACF of each jackknife sample.


\subsection{The RASS-AGN Auto-correlation Function}

\begin{figure*}
  \centering
 \resizebox{\hsize}{!}{ 
  \includegraphics[bbllx=80,bblly=381,bburx=528,bbury=660]{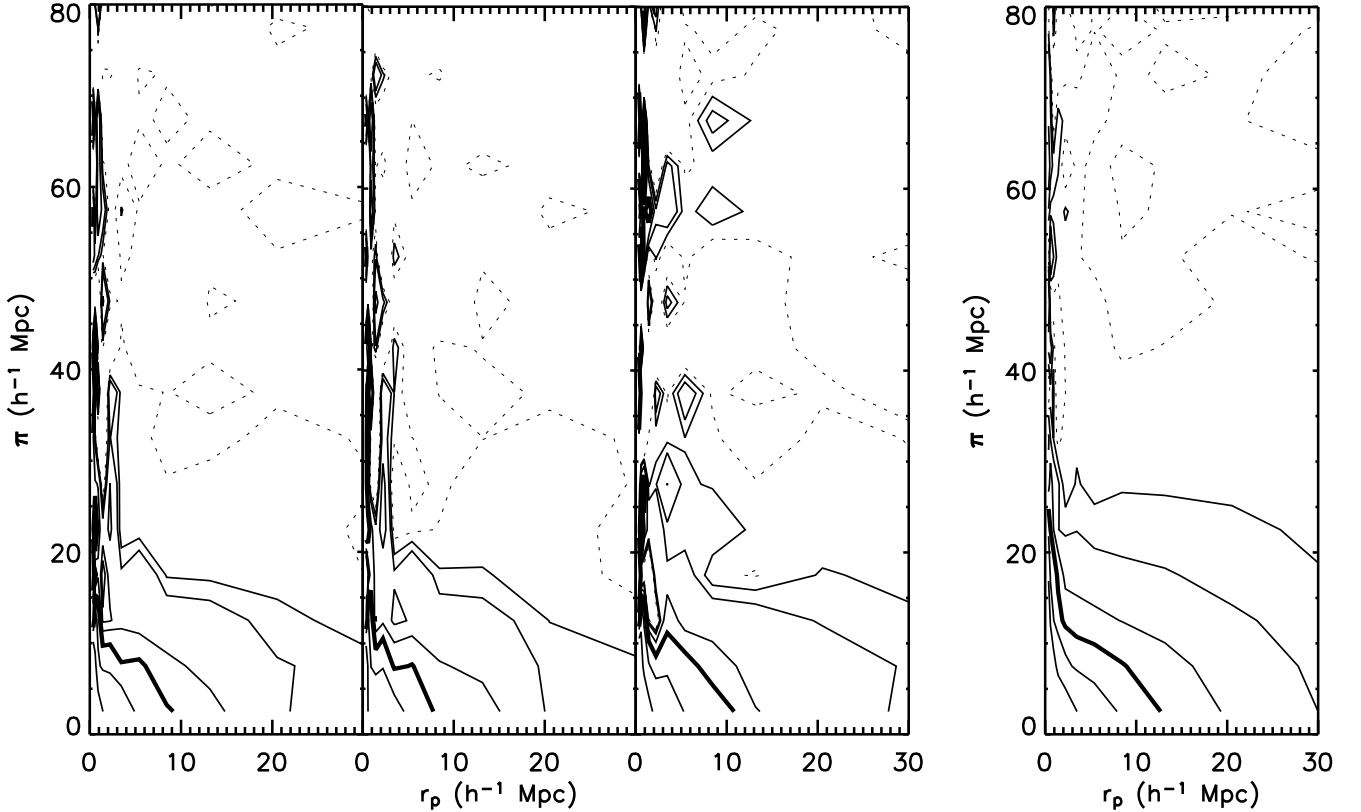}} 
      \caption{Contour plot of the CCF between RASS-AGNs 
               and LRGs for the total sample, the low $L_{\rm X}$ AGN sample, the 
               high $L_{\rm X}$ AGN sample, as well as the LRG ACF 
                (left to right). Contour lines show constant correlation strength 
               for the 2D correlation function $\xi(r_P,\pi )$. The data 
               are not smoothed. The contour levels are 0.0 (dotted line), 0.1, 
               0.2, 0.5, 1.0 (thick solid line), 2.0, and 5.0.}
         \label{contour_ccf}
\end{figure*}

We compute the CCF between RASS-AGNs and LRGs for the total sample, the 
low $L_{\rm X}$ sample, and the high $L_{\rm X}$ sample (see Table 2), as well as the ACF 
of the LRGs. We measure $r_P$ in a range of 0.3-40 $h^{-1}$ Mpc in 11 bins in a logarithmic 
scale, while $\pi$ is computed in steps of 5 $h^{-1}$ Mpc in a range of $\pi=0-200$ 
$h^{-1}$ Mpc. 
The resulting $\xi(r_P,\pi )$ are shown in Figure~\ref{contour_ccf} for $r_p=0-30$ $h^{-1}$ Mpc and
$\pi=0-80$ $h^{-1}$ Mpc. 
Note the flattened contour at $\pi \sim 40$ $h^{-1}$ Mpc in the LRG ACF. This 
is the first direct observation of the coherent infall for LRGs as expected by the Kaiser effect
(\citealt{kaiser_1987}). 
 
Although Equation~(\ref{pi_integration}) requires an integration over $\pi$ to infinity, in practice 
an upper bound of integration ($\pi_{\rm max}$) is used:
\begin{eqnarray}
   w_p(r_p) &=& 2\int_0^{\pi_{\rm max}}\!d\pi\,\xi(r_p,\pi)\ .
\end{eqnarray}
The value of $\pi_{\rm max}$ has to be large enough to 
include most correlated pairs and give a stable solution, but not be so large as to unnecessarily
increase the noise in the measurement.
To determine the appropriated $\pi_{\rm max}$  values for our correlation functions,
we determined the correlation length $r_{\rm 0}$  
for a set of $\pi_{\rm max}$ values by fitting $w_p(r_p)$ 
with a fixed $\gamma = 1.9$ over a $r_p$ range of 0.3-40 $h^{-1}$ Mpc. 

\begin{figure}
  \centering
 \resizebox{\hsize}{!}{ 
  \includegraphics[bbllx=85,bblly=369,bburx=540,bbury=700]{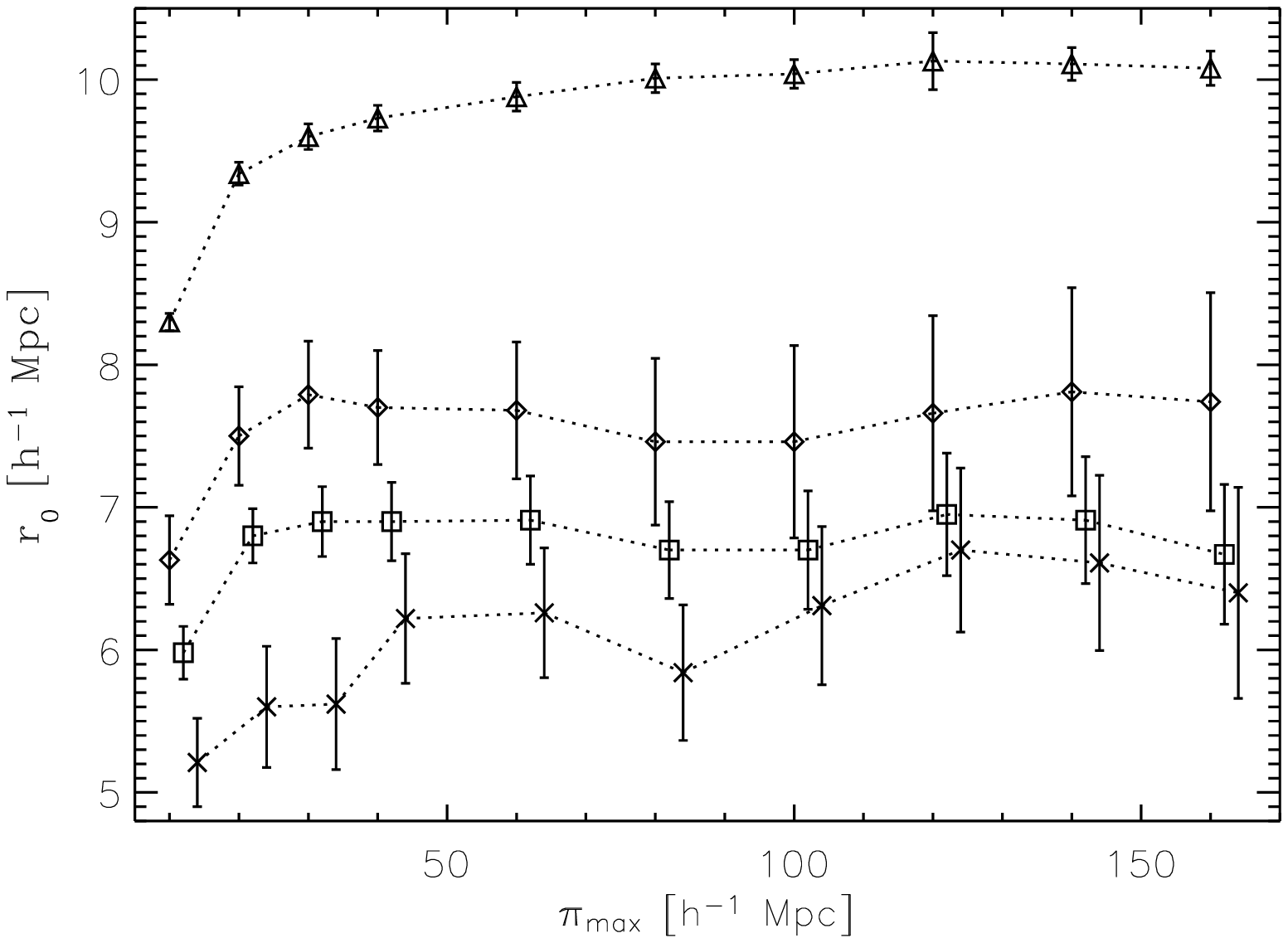}} 
      \caption{Correlation length $r_{\rm 0}$ vs. $\pi_{\rm max}$ using Equation~(\ref{pi_integration}) for 
        the LRG ACF (triangles), the high $L_{\rm X}$ RASS-AGN--LRG CCF
               (diamonds), the total RASS-AGN--LRG CCF (boxes; for illustration 
               purposes, this has been shifted +2.0 in $\pi_{\rm max}$ direction), 
               the low $L_{\rm X}$ 
               RASS-AGN--LRG CCF (crosses; shifted by +4.0 in $\pi_{\rm max}$ direction).}
              \label{rc_pimax}
\end{figure}

Figure~\ref{rc_pimax} shows that the LRG ACF saturates at $\pi_{\rm max}=80$ $h^{-1}$ Mpc. 
The changes in the correlation lengths above this value are well within the uncertainties. 
Therefore, 
as in \cite{zehavi_eisenstein_2005}, we use an upper bound of the LRG ACF integration of 
$\pi_{\rm max}=80$ $h^{-1}$ Mpc. Table~\ref{table:lrg} shows the values 
 of $r_{\rm 0}$ and $\gamma$ for a power law fit in a range of 0.3-40 $h^{-1}$ Mpc 
(as used in \citealt{zehavi_eisenstein_2005}). Both results well agree within their uncertainties. 

\begin{deluxetable}{lcc}
\tabletypesize{\normalsize}
\tablecaption{Results of Power Law Fits to the LRG ACFs and RASS-AGN--LRG CCFs \label{table:lrg}}
\tablewidth{0pt}
\tablehead{
\colhead{Sample} & \colhead{$r_{\rm 0}$ ($h^{-1}$ Mpc)} & \colhead{$\gamma$}}
\startdata
Our LRG sample   &   9.68$^{+0.14}_{-0.14}$ & 1.96$^{+0.02}_{-0.02}$ \\
Zehavi subsample1    &   9.80$\pm$0.20        & 1.94$\pm$0.02        \\\hline    
Total RASS-AGN sample           &   6.93$^{+0.27}_{-0.28}$ & 1.86$^{+0.04}_{-0.04}$\\
Low $L_{\rm X}$ RASS-AGN sample  &   6.12$^{+0.50}_{-0.53}$ & 1.94$^{+0.10}_{-0.08}$\\
High $L_{\rm X}$ RASS-AGN sample &   7.74$^{+0.40}_{-0.43}$ & 1.92$^{+0.08}_{-0.08}$
\enddata
\tablecomments{Values of $r_{\rm 0}$ and $\gamma$ are obtained from a power law fit to 
  $w_{\rm p}(r_{\rm p})$ over the range $r_p$=0.3-40 
               $h^{-1}$ Mpc for all samples using the full error covariance matrix. For the LRG ACFs, 
  \cite{zehavi_eisenstein_2005} and we used a $\pi_{\rm max}=80$ $h^{-1}$ Mpc, while for all CCFs 
  $\pi_{\rm max}=40$ $h^{-1}$ Mpc was applied.}
\end{deluxetable}

The CCFs between the different RASS-AGN samples and the LRGs saturate 
at $\pi_{\rm max}=40$ $h^{-1}$ Mpc (Figure~\ref{rc_pimax}). At higher values of $\pi_{\rm max}$, the 
signal-to-noise ratio degrades and no significant change in $r_{\rm 0}$ occurs. Therefore, we 
use $\pi_{\rm max}=40$ $h^{-1}$ Mpc as an upper bound of integration for all AGN--LRG CCFs. 
The difference in the measured LRG ACF using $\pi_{\rm max}=40$ $h^{-1}$ Mpc 
and $\pi_{\rm max}=80$ $h^{-1}$ Mpc is only 3\%. We expect that the growth of the CCF between
$\pi_{\rm max}=40$ $h^{-1}$ Mpc and $\pi_{\rm max}=80$ $h^{-1}$ Mpc is about the same order. Since this 
is much smaller than the errors in the CCF, a use of  $\pi_{\rm max}=40$ $h^{-1}$ Mpc for the CCF 
is reasonable. The correlation length of the CCFs of the different RASS-AGN samples and the LRGs 
is given in Table~\ref{table:lrg}.

As the area covered by SDSS DR4+ is not contiguous (see Figure~\ref{LRG_XAGN}), we 
computed the CCF for different subsamples of the SDSS DR4+, to check that there were no
biases introduced by using non-contiguous regions of the sky.
Excluding survey areas with R.A. $ <70\,^{\circ}$ and R.A. $ >300\,^{\circ}$, the isolated area around R.A. $ \sim 260\,^{\circ}$
and decl. $ \sim 60\,^{\circ}$, the somewhat patchy areas at 190 $\,^{\circ} < $ R.A. $ <250\,^{\circ}$ and 0$\,^{\circ} < $ decl. $ <35\,^{\circ}$, 
and combinations thereof, results in measurements of the CCFs that all agree well within their uncertainties. 
Changing the step size of $\pi$ from 5 to 2.5 $h^{-1}$ Mpc alters $w_{\rm p}(r_{\rm p})$ by 
a negligible amount. 

\begin{figure}
  \centering
 \resizebox{\hsize}{!}{ 
  \includegraphics[bbllx=67,bblly=369,bburx=540,bbury=700]{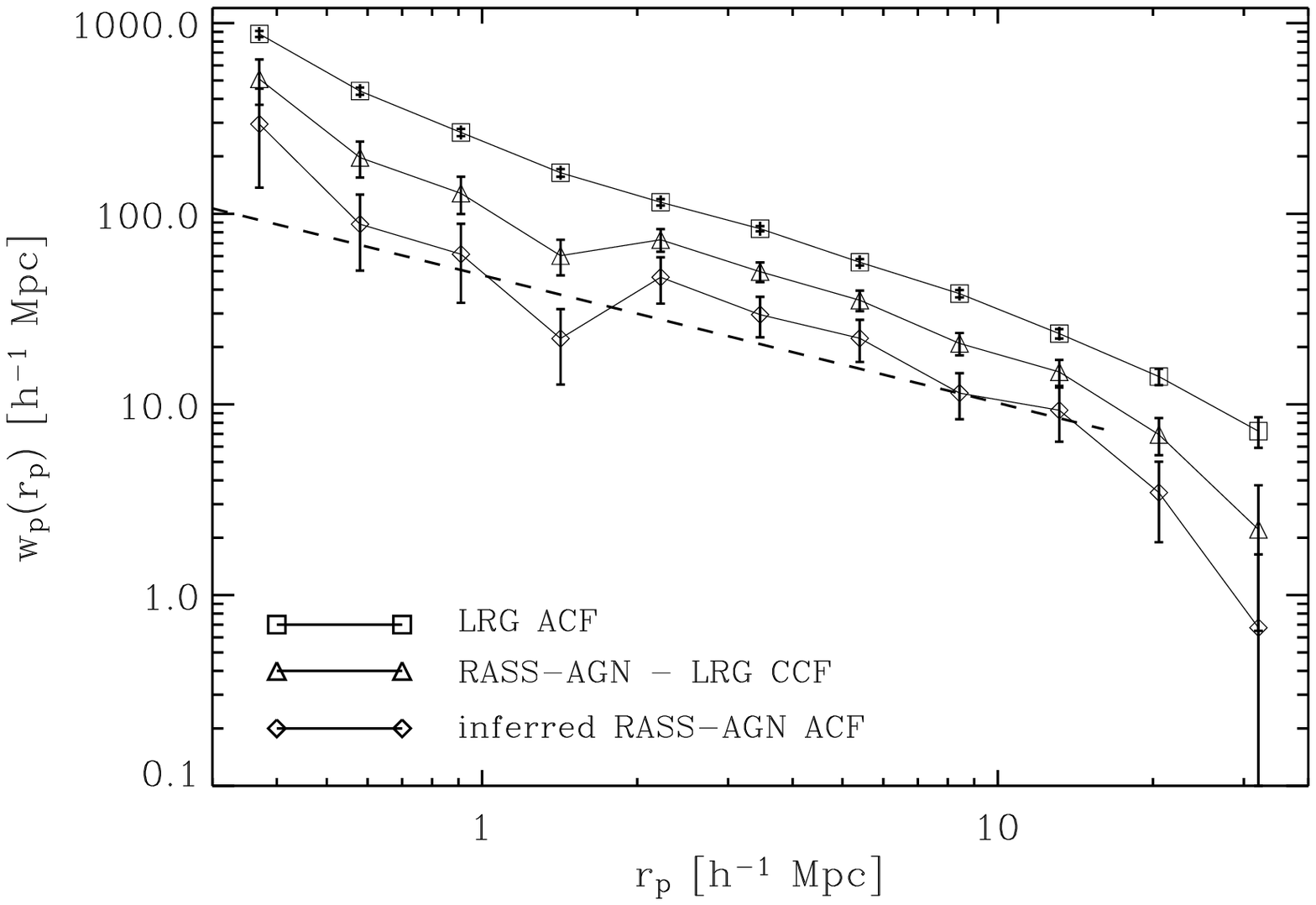}}
      \caption{Projected LRG ACF (boxes), RASS-AGN--LRG CCF (triangles), and the inferred
        RASS-AGN ACF (diamonds).
               For all data points, we show the 1$\sigma$ uncertainties. The obtained 
               best power law fit of the RASS-AGN ACF, using the covariance matrix and fitting
               over $r_p=0.3-15$ $h^{-1}$ Mpc, is shown as a dashed line. }
              \label{acf_ccf}
\end{figure}

\begin{figure}
  \centering
 \resizebox{\hsize}{!}{ 
  \includegraphics[bbllx=56,bblly=369,bburx=540,bbury=700]{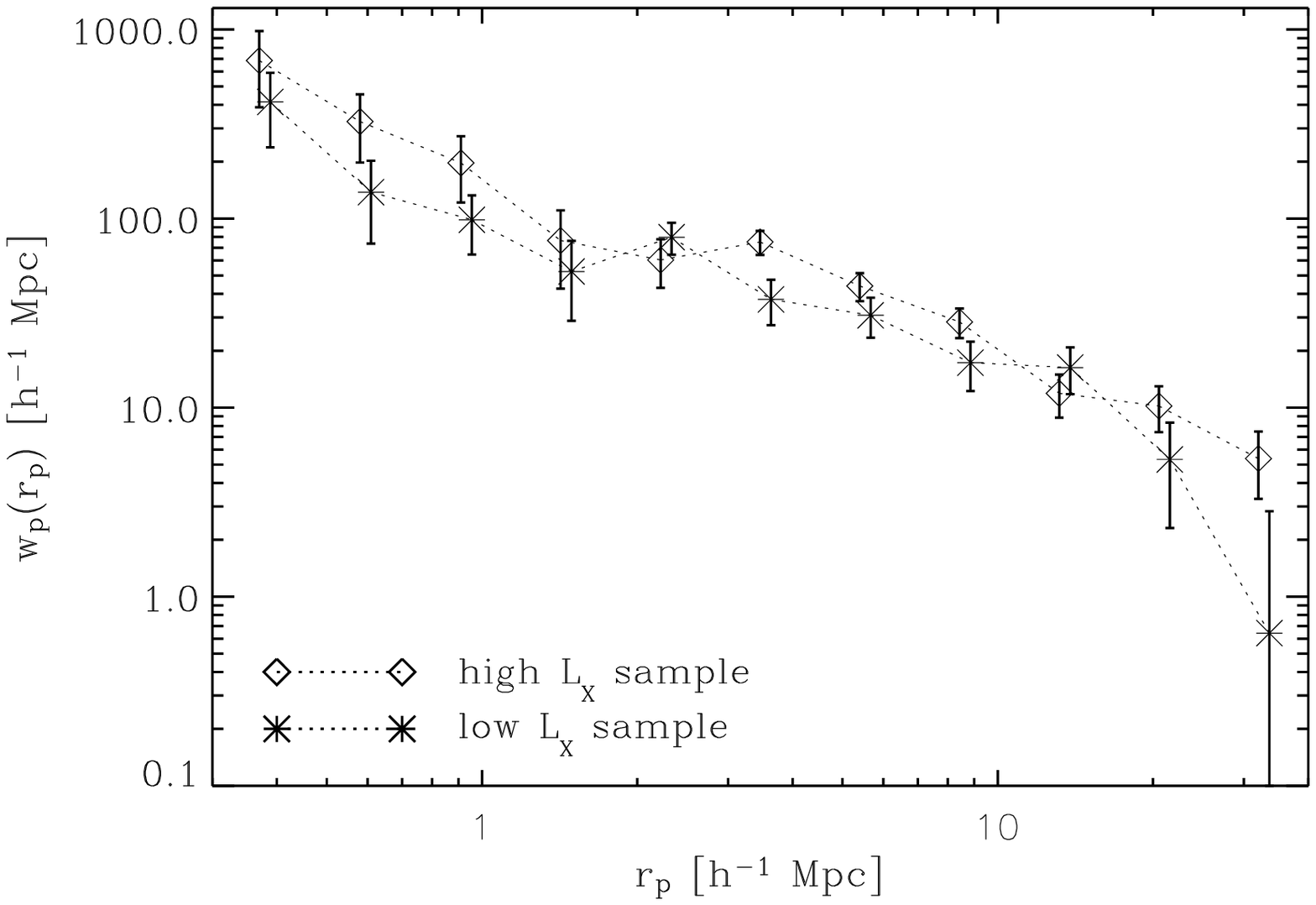}}
      \caption{CCFs of the low (stars) and high (diamonds) 
               $L_{\rm X}$ RASS-AGN samples are shown with their corresponding uncertainties.}
              \label{ccf_low_high}
\end{figure}

Instead of using the derived values of $r_0$ from the power law fits of the LRG ACF and AGN--LRG 
CCFs to compute the RASS-AGN ACF, we use Equation~(\ref{acf_rassagn}) and use the full $w_p(r_p)$ functions.
Figure~\ref{acf_ccf} shows $w_p(r_p)$ for the RASS-AGN ACF, the LRG ACF, 
and the RASS-AGN--LRG CCF. Figure~\ref{ccf_low_high} shows the CCFs for the low and high 
$L_{\rm X}$ RASS-AGN samples.

\begin{figure}
  \centering
 \resizebox{\hsize}{!}{ 
  \includegraphics[bbllx=67,bblly=369,bburx=540,bbury=700]{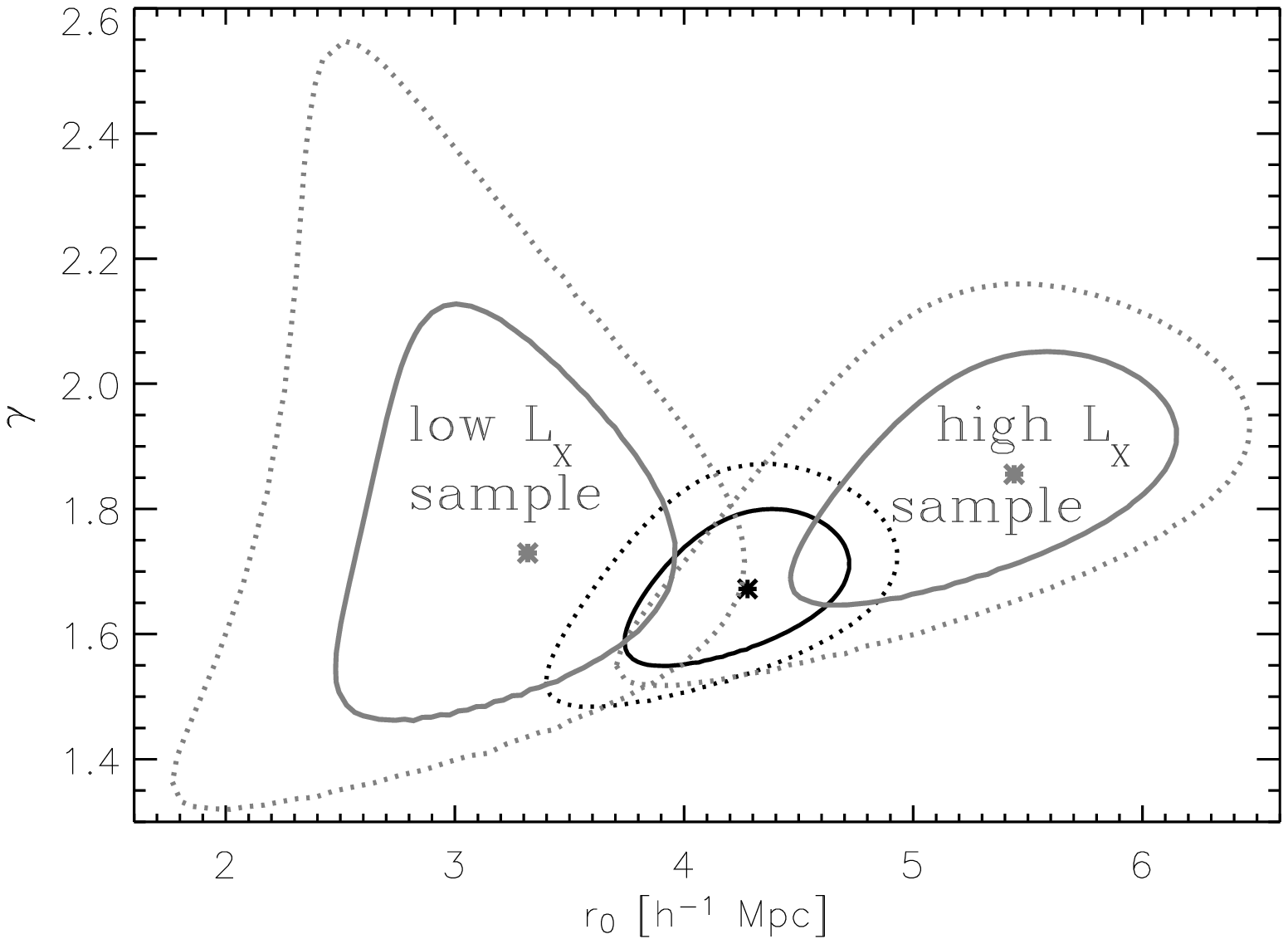}}
      \caption{Probability contours for the power law normalization $r_{\rm 0}$
               and slope $\gamma$ of the total RASS-AGN sample (black, central contour), 
               low $L_{\rm X}$ RASS-AGN sample (gray, left contour), and 
               high $L_{\rm X}$ RASS-AGN sample (gray, right contour). 
               The solid contours represent the 1$\sigma$  confidence intervals (68.3\%) 
               for a one-parameter fit based on a correlated $\chi^2= \chi^2_{\rm min} + 1.0$ 
               (Equation~(\ref{correlated_chi})), while the dotted lines illustrate the 
               corresponding intervals for a two-parameter fit 
               (1$\sigma$, correlated $\chi^2= \chi^2_{\rm min} + 2.3$).}
              \label{fit_contours}
\end{figure}

We fit power laws to the ACFs of the different RASS-AGN 
samples. The fit uses the covariance matrix and 
minimizes the correlated $\chi^2$ values according to 
\begin{eqnarray}
 \label{correlated_chi}
 \chi^2 = \sum_{i=1}^{N_{\rm bins}}\sum_{j=1}^{N_{\rm bins}}\bigg(w_p(r_{p,i})-w_p^{\rm model}(r_{p,i})\bigg) \nonumber \\
           \times\,M_{ij}^{-1} \bigg(w_p(r_{p,j})-w_p^{\rm model}(r_{p,j})\bigg)
\end{eqnarray}
We only fit the data in a range $r_p=$0.3-15 $h^{-1}$ Mpc, as the clustering signal above 
$r_p=$15 $h^{-1}$ Mpc is not well-constrained for the low $L_{\rm X}$ RASS-AGN sample. The upper end of 
$r_p$ has also been chosen because we will later convert the fit results into  $\sigma_{\rm 8,AGN}$
which involves only the pairs within 16 $h^{-1}$ Mpc. Contour plots of the resulting values of 
$r_{\rm 0}$ and $\gamma$ are shown in Figure~\ref{fit_contours} for the different RASS-AGN samples.
The derived best-fit values, as well as the best-fit $r_{\rm 0}$ values with a 
fixed power law slope of $\gamma=1.9$, are given in Table~\ref{xagn_acf}. Based on the error on 
$r_{\rm 0}$ for a fixed $\gamma=1.9$, we estimate the clustering signal to be detected at 
a $\sim$11$\sigma$, $\sim$5$\sigma$, and $\sim$8$\sigma$ level for the total, the low $L_{\rm X}$, and 
the high $L_{\rm X}$ RASS-AGN sample, respectively. The difference in the clustering signal between the 
low $L_{\rm X}$, and the high $L_{\rm X}$ RASS-AGN sample is detected at the $\sim$2.5$\sigma$ level.

\begin{deluxetable*}{lcccccc}
\tabletypesize{\normalsize}
\tablecaption{Power Law Fits to the RASS-AGN ACF.\label{xagn_acf}}
\tablewidth{0pt}
\tablehead{
\colhead{Sample} & \colhead{$r_{\rm 0}$}   & \colhead{}         &\colhead{$r_{\rm c,\gamma =1.9}$} &\colhead{}                      &\colhead{$b(z)$}                                      &\colhead{log $M_{\rm DMH}$}\\
\colhead{Name} & \colhead{($h^{-1}$ Mpc)} & \colhead{$\gamma$} &\colhead{($h^{-1}$ Mpc)       } &\colhead{$\sigma_{\rm 8,AGN}(z)$} &\colhead{($\sigma_{\rm 8,AGN}(z)/\sigma_8(z)$)}   &\colhead{($h^{-1}$ $M_{\odot}$)}}
\startdata
Total RASS-AGN sample           &   4.28$^{+0.44}_{-0.54}$ & 1.67$^{+0.13}_{-0.12}$& 4.32$^{+0.37}_{-0.41}$ & 0.77$^{+0.07}_{-0.08}$ &1.11$^{+0.10}_{-0.12}$ &  $12.58^{+0.20}_{-0.33}$\\
Low $L_{\rm X}$ RASS-AGN sample  &   3.32$^{+0.64}_{-0.83}$ & 1.73$^{+0.40}_{-0.27}$& 3.26$^{+0.58}_{-0.69}$ & 0.62$^{+0.10}_{-0.12}$ &0.88$^{+0.14}_{-0.17}$ &  11.83$^{+0.55}_{-\infty}$\\
High $L_{\rm X}$ RASS-AGN sample &   5.44$^{+0.71}_{-0.98}$ & 1.86$^{+0.20}_{-0.21}$& 5.52$^{+0.64}_{-0.71}$ & 0.98$^{+0.15}_{-0.18}$ &1.44$^{+0.22}_{-0.27}$ &  13.10$^{+0.24}_{-0.43}$
\tablecomments{Values are obtained from a fit to $w_{\rm p}(r_{\rm p})$ in a range of 0.3-15 $h^{-1}$ Mpc for all samples using the full 
               error covariance matrix.}
\end{deluxetable*}

The clustering strength is commonly expressed in terms of the rms 
fluctuation of the density distribution over the sphere with a comoving 
radius of 8 $h^{-1}$ Mpc. For a power law correlation function, 
this value can be calculated by (\citealt{miyaji_zamorani_2007}; Section 59 of \citealt{peebles_1980})
\begin{eqnarray}
  (\sigma_{\rm 8,AGN})^2=J_2(\gamma)\left(\frac{r_{\rm 0}}{8\,h^{-1}\,{\rm Mpc}}\right)^{\gamma},
\end{eqnarray} 
where $J_2(\gamma)= 72/\left[(3-\gamma)(4-\gamma)(6-\gamma)2^{\gamma}\right]$.
The uncertainty of $\sigma_{\rm 8,AGN}$ is computed from the $r_{\rm 0}$ versus $\gamma$ 
confidence contour of the one-parameter fit based on a correlated $\chi^2= \chi^2_{\rm min} + 1.0$ 
(Figure~\ref{fit_contours}).

Based on $\sigma_{\rm 8,AGN}$, we further calculate the RASS-AGN bias parameter
$b=\sigma_{\rm 8,AGN}(z)/\sigma_8(z)$. This quantity allows us to compare the 
observed AGN clustering to the underlying mass distribution from linear growth theory 
(\citealt{hamilton_2001}). We use a normalization to a value of $\sigma_8(z=0)=0.8$
which is consistent with the {\em Wilkinson Microwave Anisotropy Probe (WMAP)} DR5 for a $\Lambda$CDM+SZ+LENS 
model\footnote{\tt http://lambda.gsfc.nasa.gov/product/map/dr3/params/lcdm\_sz\_lens\_wmap5.cfm}.
The errors on $b$ are derived from the standard deviation of $\sigma_{\rm 8, AGN}$. 

Because we measure the CCF to infer the ACF, the resulting effective redshift distribution 
for the clustering signal
is determined by both the redshift distribution of the LRG sample and the RASS-AGN sample: 
$N_{\rm CCF}(z) = N_{\rm LRG}(z)*N_{\rm RASS-AGN}(z)$. The median redshift of $N_{\rm CCF}(z)$
is $\overline{z_{\rm CCF}}=0.27$, 0.24, 0.31 for the total, the low $L_{\rm X}$, and the high $L_{\rm X}$ RASS-AGN 
samples, respectively. The difference compared to the mean redshift of the different RASS-AGN samples (see 
Table~\ref{xagn_samples}) is at most $\Delta z=0.03$. Our measurements 
of $\sigma_{\rm 8, AGN}$ and $b(z)$ (using $\overline{z_{\rm CCF}}$) for all RASS-AGN samples are listed in 
Table~\ref{xagn_acf}. 

{\it If} RASS-AGNs are hosted by typical $L^{\ast}$ galaxies, the fraction of $L^{\ast}$ galaxies 
hosting RASS-AGNs can be calculated using the observed number density of both host galaxies and RASS-AGNs. 
\cite{blanton_hogg_2003} measure the number density  of  $L^{\ast}$ galaxies at $z=0.1$ 
to be $\phi^{\ast}_{r\rm{-band}} = 0.0149\pm 0.0004$ $h^3$ Mpc$^{-3}$.
Using the observed number density in the total RASS-AGN sample (Table~\ref{xagn_samples}), 
we find that only $\sim$0.4\% of $L^{\ast}$ galaxies could harbor RASS-AGNs.

\subsection{Host Dark Matter Halo Mass}
Our clustering results can be used to estimate the ``typical'' dark matter halo mass $M_{\rm DMH}$ hosting our 
different AGN samples.
The dark matter halo mass is reflected by the bias parameter $b$, which reflects the clustering amplitude relative to the underlying dark 
matter distribution. Using Equation~8 of \cite{sheth_mo_2001}, we compute the expected large-scale Eulerian 
bias factor for different dark matter halo masses. The required ratio of the critical overdensity to the rms 
fluctuation on a given size and mass is calculated by $\nu = \delta_{\rm cr}/\sigma(M,z)$ 
(\citealt{sheth_mo_2001}). Assuming $\delta_{\rm cr}\approx 1.69$, we compute $\sigma(M,z)$ using 
Equations~(A8), (A9), and (A10) from \cite{bosch_2002}; this approach was also used in \cite{hickox_jones_2009}. 
The typical dark matter halo mass for the total RASS-AGN and the high $L_{\rm X}$ RASS-AGN sample is 
found to be log $(M_{\rm DMH}/(h^{-1}\,M_\odot)) =12.58^{+0.20}_{-0.33}$, 13.10$^{+0.24}_{-0.43}$,
respectively. 
For the low $L_{\rm X}$ RASS-AGN sample (log $(M_{\rm DMH}/(h^{-1}\,M_\odot)) =11.83^{+0.55}$) we are 
not able to constrain a lower limit on the dark matter halo mass. This is because 
our lower limit corresponds to $b=0.71$ and the minimum $b$ value derived from \cite{sheth_mo_2001} 
at $z=0.24$ is $b_{\rm min}=0.72$ (at $M_{\rm DMH} \sim 10^{9.3}$ $h^{-1}$ $M_\odot$).


\section{Discussion}

\subsection{Comparison to Other X-ray Clustering Measurements}
 
Our findings can be directly compared to previous attempts to measure the spatial ACF 
(using spectroscopic redshifts) of low-$z$ RASS-based AGNs at similar X-ray luminosities. 
\cite{grazian_negrello_2004} found a redshift-space clustering length of 
$s_{\rm 0}=8.64^{+2.00}_{-2.08}$ $h^{-1}$ Mpc 
in the ASIAGO-ESO/RASS QSO survey when fitting their correlation function 
with $\gamma$ fixed at 1.56. They did not publish results for fits of $r_{\rm 0}$ and $\gamma$. 
A second low-$z$ RASS-AGN clustering study with very similar AGN properties was conducted by 
\cite{mullis_henry_2004} in the {\em ROSAT} North Ecliptic Pole survey. Their best-fit values 
are $r_{\rm 0}=7.5^{+2.7}_{-4.2}$ $h^{-1}$ Mpc and $\gamma=1.85^{+1.90}_{-0.80}$. 
Our inferred ACF of the total RASS-AGN sample yield 
$r_{\rm 0}=4.28^{+0.44}_{-0.54}$ $h^{-1}$ Mpc and $\gamma=1.67^{+0.13}_{-0.12}$. 
Both studies used Poisson errors which underestimate systematic effects. 
Using the largest sample of X-ray selected AGNs ever applied to clustering measurements, and by measuring 
the cross-correlation with more numerous galaxies, we reach significantly lower uncertainties compared 
to the previous studies and detect a clustering signal at the $\sim$11$\sigma$ level. We also find 
a much smaller correlation length. \cite{grazian_negrello_2004} and \cite{mullis_henry_2004}
derive their results from fitting the ACF at much larger separations ($s$, $r_P$ values)
than we do. However, considering their larger uncertainties, their correlation lengths differ from ours  
by only 1$\sigma$--2$\sigma$.
The low number of X-ray selected AGNs used in both studies did not allow them to 
split their samples in bins of either luminosity or redshift. 

\begin{deluxetable*}{lcccccccccc}
\tabletypesize{\normalsize}
\tablecaption{Comparison of Published Real-space Clustering Measurements of X-ray Selected AGNs and Optically Selected AGNs, as well as Red and Blue Galaxies.\label{table:study}}
\tablewidth{0pt}
\tablehead{
\colhead{}               & \colhead{$r_{\rm 0}$}     & \colhead{}          &\colhead{Fitted}   &\colhead{Area}      &\colhead{Object}  &\colhead{$L$,$M$}   &\colhead{Band}         &\colhead{}            &\colhead{Bias}  &\colhead{}\\
\colhead{Sample}         & \colhead{($h^{-1}$ Mpc)} & \colhead{$\gamma$}  &\colhead{Range}    &\colhead{(deg$^2$)}  &\colhead{Number}  &\colhead{(erg/s),(mag)}               &\colhead{(keV),filter} &\colhead{$z$}  &\colhead{$b$} &\colhead{Ref.}}
\startdata
RASS   & 4.28$^{+0.44}_{-0.54}$     & 1.67$^{+0.13}_{-0.12}$ & 0.3-15                & 5468        &1552        & $1.4\times 10^{44}$          & 0.1-2.4&0.27&      1.11$^{+0.10}_{-0.12}$ & This work\\
Low-$L$ RASS&3.32$^{+0.64}_{-0.83}$ & 1.73$^{+0.40}_{-0.27}$ & 0.3-15                & 5468        &990         & $9.8\times 10^{43}$          & 0.1-2.4&0.24&      0.88$^{+0.14}_{-0.17}$ & This work\\
High-$L$ RASS&5.44$^{+0.71}_{-0.98}$& 1.86$^{+0.20}_{-0.21}$ & 0.3-15                & 5468        &562         & $3.4\times 10^{44}$          & 0.1-2.4&0.31&      1.44$^{+0.22}_{-0.27}$& This work\\\hline
\multicolumn{11}{c}{X-ray Selected AGN Clustering Measurements}\\
NEP    & 7.5$^{+2.7}_{-4.2}$        & 1.85$^{+1.90}_{-0.80}$ & 5-60$^{\rm P}$         & 80.7        & 219        & $9.2\times 10^{43}$          &0.5-2.0&0.22$^{E}$& 1.83$^{+1.88}_{-0.61}$ & Mu04\\
CDFN   & $5.5\pm 0.6$             & $1.50 \pm 0.12$      & 0.2-10$^{\rm P}$       & 0.13        & 160        & $1\times 10^{43}$            &0.5-10 &0.96&       1.87$^{+0.14}_{-0.16}$ & Gi05\\
CDFS   & $10.3\pm 1.7$            & $1.33 \pm 0.14$      & 0.2-10$^{\rm P}$       & 0.1         & 97         & $1.6\times 10^{43}$          &0.5-10 &0.84&       2.64$^{+0.29}_{-0.30}$ & Gi05\\
CLASXS & 8.1$^{+1.2}_{-2.2}$        & $2.1 \pm 0.5$        & 1-30                  & 0.4         & 233        & $9\times 10^{43}$            & 2-8   &1.2 &       3.58$^{+2.49}_{-1.38}$ & Ya06\\
CDFN   & 5.8$^{+1.0}_{-1.5}$        & 1.38$^{+0.12}_{-0.14}$ & 0.2-15                & 0.13        & 252        & $3\times 10^{42}$            & 2-8   &0.8 &       1.77$^{+0.80}_{-0.15}$ & Ya06\\              
AEGIS  & $5.95\pm 0.90$           & $1.66 \pm 0.22$      & 0.1-8                 & 0.63        & 113        & $6.3 \times 10^{42}$         &2-10   &0.90&       1.97$^{+0.26}_{-0.25}$ & Co09\\
COSMOS & $8.65^{+0.41}_{-0.48}$     & $1.88^{+0.06}_{-0.07}$ & 0.3-40$^{\rm P}$       & 1.96        & 538        & $6.3 \times 10^{43}$         &0.5-10 &0.98&       3.08$^{+0.14}_{-0.14}$ & Gi09\\\hline 
\multicolumn{11}{c}{Optically Selected AGN Clustering Measurements}\\
2QZ    & $4.8^{+0.9}_{-1.5}$        & $1.53^{+0.19}_{-0.21}$ & 0.8-20                & $\sim$700   & 13989      & -23.82                      & $b_{\rm J}$ &1.47$^{E}$&2.07$^{+0.35}_{-0.44}$  & Po04\\
2QZ-subs1&$5.4^{+0.9}_{-1.3}$       & $2.02^{+0.36}_{-0.33}$ & 2-20                  & $\sim$700   & 4928       & -23.13                      & $b_{\rm J}$ &1.06$^{E}$&2.14$^{+0.71}_{-0.55}$  & Po04\\
2QZ-subs2&$4.3^{+1.8}_{-2.0}$       & $1.49^{+0.32}_{-0.35}$ & 2-20                  & $\sim$700   & 4737       & -23.84                      & $b_{\rm J}$ &1.51$^{E}$&1.93$^{+0.73}_{-0.90}$  & Po04\\
2QZ-subs3&$7.6^{+1.2}_{-2.1}$       & $1.79^{+0.25}_{-0.29}$ & 2-20                  & $\sim$700   & 4324       & -24.30                      & $b_{\rm J}$ &1.89$^{E}$&3.71$^{+0.97}_{-1.03}$  & Po04\\
DEEP2    &$3.1\pm 0.6$          & $1.8$(fixed)       & 0.1-10                & 3           & 52         & -23.0                       & $B$        & 0.99     &1.18$^{+0.20}_{-0.20}$  & Co07\\
SDSS      &$5.45^{+0.35}_{-0.45}$   &$1.90^{+0.04}_{-0.03}$  & 1-130                 & $\sim$4000  & 30239      & $\sim$-25.8                 & $i$        &$\langle$1.27$\rangle$&2.26$^{+0.14}_{-0.18}$  & Ro09\\\hline
\multicolumn{11}{c}{Galaxy Clustering Measurements}\\
red-2dF   &$6.10 \pm 0.34$        &$1.95 \pm 0.03$       & 0.2-20                & $\sim$700   & 36318      & 1.26                        & $L^{\ast}$  &0.11      &1.41$^{+0.08}_{-0.08}$  & Ma03\\       
blue-2dF  &$3.67 \pm 0.30$        &$1.60 \pm 0.04$       & 0.2-20                & $\sim$700   & 60473      & 0.95                        & $L^{\ast}$  &0.11      &0.85$^{+0.05}_{-0.05}$  & Ma03\\
red-SDSS  &$5.67 \pm 0.37$        &$2.08 \pm 0.05$       & 0.1-10                & 2497        & 5804       & $\langle$-19.5$\rangle$                 & $r$        &$\langle$0.05$\rangle$ &1.41$^{+0.09}_{-0.09}$ & Ze05b \\
blue-SDSS &$3.63 \pm 0.16$        &$1.69 \pm 0.04$       & 0.1-10                & 2497        & 8419       & $\langle$-19.5$\rangle$                 & $r$        &$\langle$0.05$\rangle$ &0.86$^{+0.03}_{-0.04}$ & Ze05b \\
LRG-SDSS  &$9.80 \pm 0.20$        &$1.94 \pm 0.02$       & 0.3-30                & $\sim$3800  & 29298      & $\langle$-21.63$\rangle$                & $g_{z=0.3}$ &$\langle$0.28$\rangle$&2.56$^{+0.06}_{-0.06}$  & Ze05a\\ 
red-AGES & $5.3\pm 0.2$           &$2.1\pm 0.1$          & 0.3-10                & 7           & 3146       & -21.3                       & $r_{z=0.1}$ &0.41      &1.59$^{+0.06}_{-0.06}$  & Hi09\\
blue-AGES & $3.8\pm 0.2$          &$1.6\pm 0.1$          & 0.3-10                & 7           & 3116       & -21.0                       & $r_{z=0.1}$ &0.38      &1.07$^{+0.05}_{-0.05}$  & Hi09\\
red-DEEP2 & $5.25\pm 0.26$        &$2.06\pm 0.04$        & 0.1-20                & 3           & 1474       & -20.70                      & $B$        &$\langle$0.82$\rangle$&1.88$^{+0.09}_{-0.09}$ & Co08\\
blue-DEEP2& $3.87\pm 0.12$        &$1.64\pm 0.05$        & 0.1-20                & 3           & 4808       & -20.48                      & $B$        &$\langle$0.90$\rangle$&1.39$^{+0.04}_{-0.04}$ & Co08\\
red-VVDS  & $3.78^{+0.70}_{-0.74}$  &$1.87^{+0.28}_{-0.22}$  & 0.1-10                & 0.5         & 355        & -19.57                      & $B_{AB}$    &0.81$^{E}$&1.31$^{+0.23}_{-0.24}$  & Me06\\
blue-VVDS & $2.49^{+0.28}_{-0.22}$  &$1.84^{+0.14}_{-0.10}$  & 0.1-10                & 0.5         & 1105       & -19.75                      & $B_{AB}$    &1.04$^{E}$&0.98$^{+0.10}_{-0.08}$  & Me06
\tablecomments{Values for the fitted range (Column 4) are in units of $h^{-1}$ Mpc. 
Values listed for $L$ and $M$ are the median values whenever these quantities were available. Otherwise, mean values are given and denoted by angled brackets ($\langle$$\rangle$).
Superscript ``P'' on the fitted range indicates that only Poisson errors have been used for the fit, which causes a significant 
underestimation of the uncertainties. All other studies used either the bootstrap or jackknife method to estimate their uncertainties.  
Superscript ``{\em E}'' shows that the effective redshift was given in the study, as opposed to the median redshift. Abbreviation: subs--subsample.  
References: Gr04--\cite{grazian_negrello_2004}; Mu04--\cite{mullis_henry_2004}; Gi05--\cite{gilli_daddi_2005}; 
Ya06--\cite{yang_mushotzky_2006}; Co09--\cite{coil_georgakakis_2009}; Gi09--\cite{gilli_zamorani_2009}; 
Hi09--\cite{hickox_jones_2009}; Po04--\cite{porciani_magliocchetti_2004}; Co07--\cite{coil_hennawi_2007}; 
Ro09--\cite{ross_shen_2009}; Ma03--\cite{madgwick_hawkins_2003}; Ze05b--\cite{zehavi_zheng_2005} 
Ze05a--\cite{zehavi_eisenstein_2005}; Co08--\cite{coil_newman_2008}; Me06--\cite{meneux_fevre_2006}.   }
\end{deluxetable*}

At higher redshift, data from more modern X-ray telescopes 
such as {\em XMM-Newton} and {\em Chandra} have been used to measure the clustering of 
X-ray selected AGNs in various deep survey fields (Table~\ref{table:study}). These surveys are 
significantly more sensitive than {\em ROSAT} surveys but have sampled smaller comoving volumes 
(due to smaller sky area coverage) and may therefore be affected by cosmic variance.
The median X-ray luminosity of these surveys is roughly 1 mag fainter than 
the X-ray luminosities detected in {\em ROSAT}-based catalogs. Furthermore, 
{\em XMM-Newton} and {\em Chandra} are sensitive at both soft energy bands (as {\em ROSAT}: 0.1-2.4 keV) 
and hard energy bands (2-10 keV). This influences the sample selection; {\em ROSAT} samples are dominated by 
X-ray unabsorbed type I AGNs, while {\em XMM-Newton} and {\em Chandra} samples consist of a nearly equal 
mix of type I AGNs and absorbed type II AGNs. A direct comparison of {\em ROSAT} and 
{\em XMM-Newton}/{\em Chandra} AGN clustering measurements is therefore challenging.

The inferred typical host dark matter mass for RASS-AGNs at $z=0.27$ is consistent 
with that found at higher redshift in CDFN and AEGIS, though other X-ray AGN clustering studies 
find significant larger values at these redshifts 
(i.e., CDFS, CLASXS, and COSMOS; see Table~\ref{table:study} for details).

One possible explanation of the differences in the clustering signals between the low and high-$z$ 
results could be the presence of a large fraction of type II AGNs in the {\em XMM-Newton} and {\em Chandra} 
samples. However, \cite{gilli_daddi_2005, gilli_zamorani_2009} measured the 
correlation strengths separately for soft and hard AGN samples, as well as for broad emission line 
versus narrow emission line AGN samples, and did not find a significant difference.  However,  larger 
samples with smaller uncertainties are needed to definitively test this question.
\cite{gilli_daddi_2005, gilli_zamorani_2009} explain the large values of $r_{\rm 0}$ in the CDFS and 
COSMOS by the cosmic variance, i.e., very prominent redshift spikes at $z\sim0.7$ and $z\sim 0.36$, 
respectively. Removing these overdensities in the total X-ray sample results in correlation lengths of  
$r_{\rm 0}=3.8^{+1.3}_{-2.7}$, 6.1$\pm 0.8$ $h^{-1}$ Mpc, respectively, similar to the values 
found here at lower redshift. This result underscores the need to include cosmic variance errors 
when using samples that do not probe a large volume. Among measurements that correctly account for cosmic variance, 
cosmic evolution and differences in X-ray luminosity may be responsible for the different clustering results 
in the literature. 
We will discuss this point in detail in Section~\ref{x-ray_dependence}.  

In Table~\ref{table:study}, we compare the main properties of the different clustering measurements 
of AGNs selected at different wavelengths (e.g., X-ray and optical) and star-forming and quiescent 
(i.e., blue and red) galaxies at similar redshifts. Table~\ref{table:study} aims to compare different studies as uniformly as possible to detect 
general trends in the clustering of AGNs to $z\sim1$.  
Therefore only studies that are based on the real-space correlation function $\xi(r)$ are included here.  
Many results, especially for optically selected AGNs, 
measure only the redshift-space correlation function, $\xi(s)$, and attempt to model redshift-space 
distortions to infer $\xi(r)$. However, $\xi(s)$ is not well-approximated by a power law and the 
derived results depend therefore on the fitted $s$ range. Additionally, systematic uncertainties in 
the modeling make direct comparisons to $\xi(r)$ measurements extremely difficult.
To facilitate comparison between samples, we recalculate the bias, $b(z)$, for each study given in Table~\ref{table:study} 
in a consistent manner.  Values of $\sigma_{\rm 8, AGN/GAL}$ are computed using confidence contours in 
$r_{\rm 0}$--$\gamma$ space (where $\chi^2 = \chi_{\rm best fit}^2 +1$) published in the literature. 
If no confidence contours are given, we use the best-fit $r_{\rm 0}$, $\gamma$ values and 
propagate their error to derive the 1$\sigma$ error on $\sigma_{\rm 8, AGN/GAL}$. We exclude papers 
that report only $b$ values but do not publish $r_{\rm 0}$ and $\gamma$ values. 
Where possible, we report the median redshift of objects in each study 
that contribute to the clustering signal. 
We use $z_{\rm eff}$ whenever available, otherwise we use the median redshift ($\overline{z}$), 
or the mean redshift ($\langle$$z$$\rangle$), in that order, to compute $\sigma_8(z)$ using our normalization 
of $\sigma_8(z=0)=0.8$ throughout. 

\begin{figure}
  \centering
 \resizebox{\hsize}{!}{ 
  \includegraphics[bbllx=56,bblly=359,bburx=545,bbury=706]{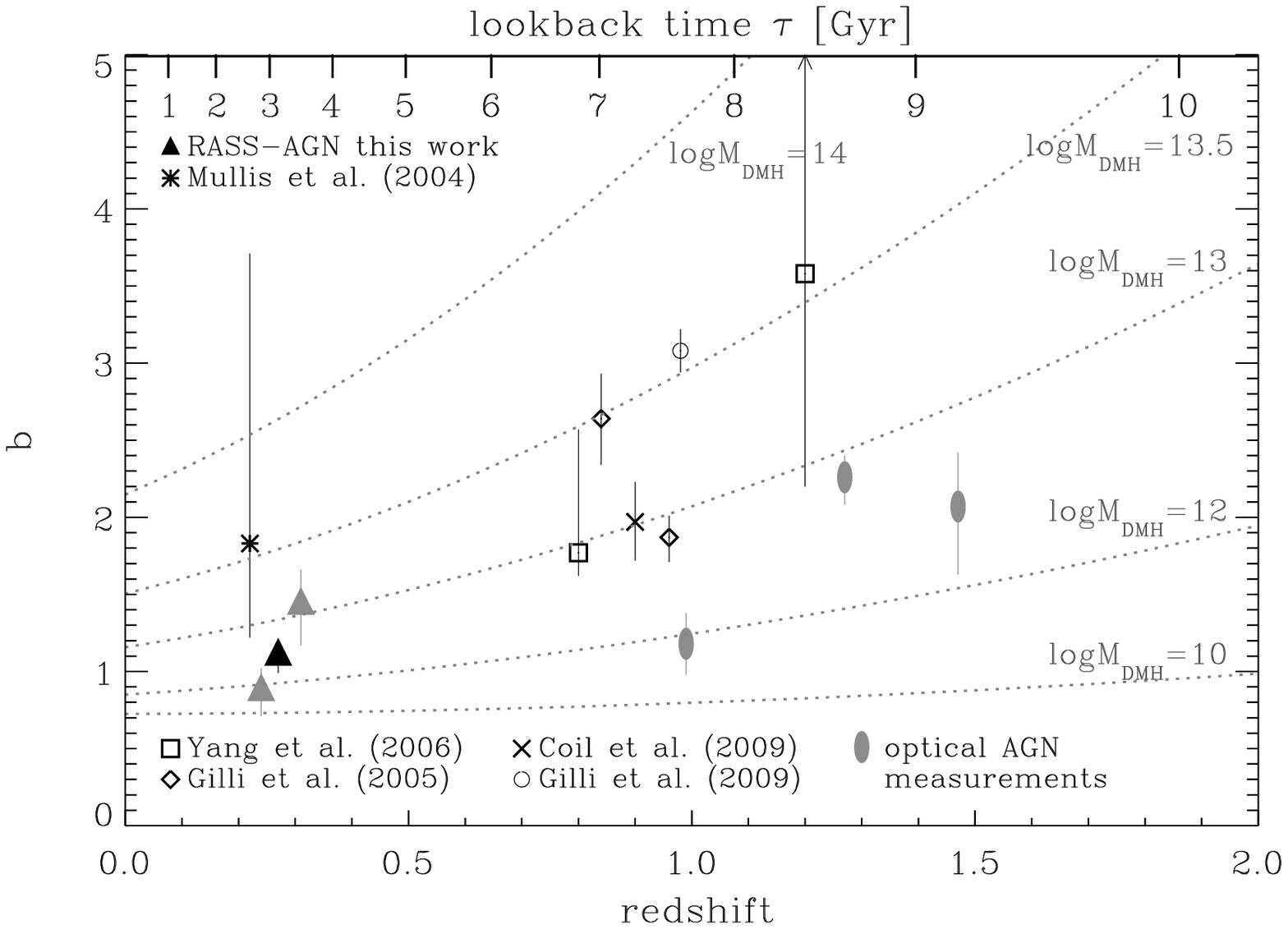}}
      \caption{Bias parameter $b_{\rm AGN}=\sigma_{8,\rm AGN}(z)/[\sigma_8D(z)]$ as a 
               function of redshift for various X-ray AGN selected samples (see the text for details). 
               Gray triangles represent luminosity-dependent subsamples 
               of the main AGN sample used here. The gray filled ellipses indicate optical 
               AGN clustering measurements (left to right: \cite{coil_hennawi_2007}; 
               \cite{ross_shen_2009}; \cite{porciani_magliocchetti_2004}). The dotted lines 
               show the expected $b(z)$ of typical dark matter halo masses $M_{\rm DMH}$ based on \cite{sheth_mo_2001}. 
               The masses are given in log $M_{\rm DMH}$ in units of $h^{-1}$ $M_\odot$.}
              \label{xagn_b}
\end{figure}

Figure~\ref{xagn_b} shows the derived bias factors $b$ of various X-ray selected AGN samples as a function 
of redshift. The results shown in this figure suggest that high redshift, low X-ray luminous AGNs 
tend to reside in higher mass dark matter halos ($M_{\rm DMH} \sim 10^{13}$ $h^{-1}$ $M_\odot$) 
than low redshift, high luminous AGNs. Broad-line X-ray and optical AGN measurements seem to show a lower 
bias values than the narrow-line and/or fainter X-ray AGN measurements. 

\begin{figure}
  \centering
 \resizebox{\hsize}{!}{ 
  \includegraphics[bbllx=56,bblly=359,bburx=545,bbury=706]{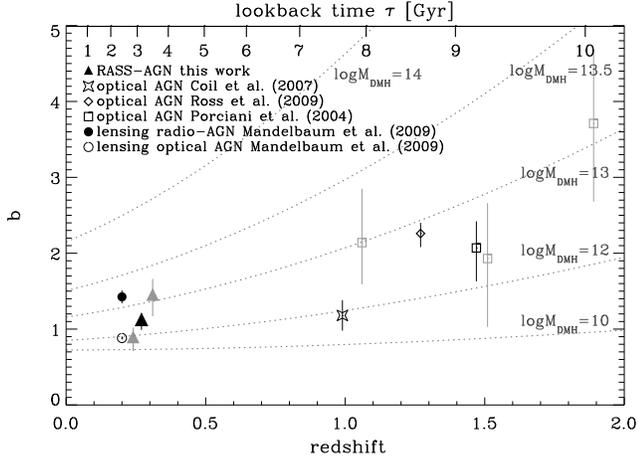}}
      \caption{Similar to Figure~\ref{xagn_b}, now comparing our results to AGN samples selected at 
               optical and radio wavelengths. Our results are shown
               at $z=0.3$; gray triangles represent 
               subsamples of the main AGN sample used here; gray boxes show subsamples of the AGN 
               sample by \cite{porciani_magliocchetti_2004}.}
              \label{oagn_b}
\end{figure}

\begin{figure}
  \centering
 \resizebox{\hsize}{!}{ 
  \includegraphics[bbllx=56,bblly=359,bburx=545,bbury=706]{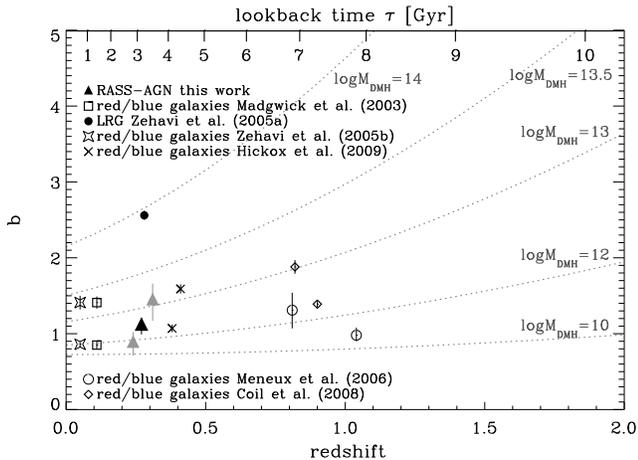}}
      \caption{Similar to Figure~\ref{xagn_b}, now comparing our results to red and
               blue galaxy samples. For all galaxy results, the data point with lower  
                $b$ value corresponds to blue, star-forming galaxies while red galaxies 
                 have higher $b$ values. }
              \label{gagn_b}
\end{figure}

\subsection{Comparison to AGN Clustering Measurements at Different Wavelengths}
The availability of large optical data sets such as SDSS, 2dF QSO Redshift Survey (2QZ), and 
2dF-SDSS LRG and QSO (2SLAQ) led to a series of large-scale clustering studies.  
However, many authors report only $\xi(s)$ (e.g., \citealt{croom_boyle_2005}; \citealt{angela_shanks_2008}),
which makes it difficult to compare their results to our $\xi(r)$-based results. 
Other studies compute the CCF 
between optically selected QSOs and LRGs (e.g., \citealt{mountrichas_sawangwit_2009}), but do not infer the 
QSO ACF.  
Here we compare only to studies that measured $\xi(r)$, listed in 
Table~\ref{table:study} (middle section). 
Most of the optically selected AGN samples contain samples that are orders of magnitude 
larger than X-ray selected samples.   
\cite{coil_hennawi_2007} use the CCF between $\sim$5000 DEEP2 galaxies and 52 QSOs to infer the QSO ACF.

AGNs selected at optical wavelengths have relatively 
low clustering scale lengths, found over a wide range of redshifts. This 
translates into low bias parameters, not much larger than what we find for our low-$z$ 
RASS-AGNs (Figure~\ref{oagn_b}). 
Radio-loud AGNs are found to be significantly more clustered than optically selected AGNs
(e.g., \citealt{magliocchetti_maddox_2004}; \citealt{hickox_jones_2009}; \citealt{mandelbaum_li_2009}).
Optically bright and radio-loud quasars are also luminous in X-rays 
(\citealt{wilkes_tananbaum_1994}) and reside in very massive dark matter halos 
(e.g., \citealt{shen_strauss_2009}).

Comparing the general trends seen in Figure~\ref{xagn_b} and \ref{oagn_b}, X-ray and optically selected 
AGNs show similar clustering strength at $z<1$. Optically selected AGNs at higher redshifts are found to 
have a lower clustering than X-ray selected AGNs. The sample of \cite{porciani_magliocchetti_2004} 
has a higher mean luminosity and higher clustering signal than the sample of \cite{coil_hennawi_2007}.
Within the errors, the clustering of optically selected AGNs 
is known to be in accordance with a redshift-independent dark matter halo mass 
of $M_{\rm DMH} \sim 10^{12}-10^{13}$ $h^{-1}$ $M_\odot$ (\citealt{porciani_magliocchetti_2004}; 
\citealt{coil_hennawi_2007}; \citealt{ross_shen_2009}). Low-luminosity AGNs ($-23 < M_{0.1_{r}}<-17$) 
in the local universe have a typical dark matter halo mass of 
$M_{\rm DMH} \sim 8 \times 10^{11}$ $h^{-1}$ $M_\odot$ (\citealt{mandelbaum_li_2009}), which is also
consistent with the $M_{\rm DMH}$ given by \cite{porciani_magliocchetti_2004} and 
\citealt{coil_hennawi_2007} for the high-redshift luminous AGNs considering the uncertainties 
in their clustering measurements.

\subsection{Luminosity Dependence of the Clustering Signal\label{x-ray_dependence}}
In the hierarchical model of structure formation, more massive galaxies should reside in more 
massive dark matter halos and therefore be more strongly clustered. More massive galaxies are 
expected to be more luminous, which should lead to a luminosity dependence of the clustering 
signal. This result has been confirmed by clustering measurements of, e.g., SDSS galaxies 
at low redshifts (\citealt{zehavi_zheng_2005}) and of, e.g., DEEP2 galaxies at $z=1$ 
(\citealt{coil_newman_2006}). Whether this relation applies also to the AGN luminosity
is unclear. The AGN luminosity depends on the SMBH mass, mass accretion rate, and radiative 
accretion efficiency. In the simple case in which all SMBHs have the same 
Eddington ratio and the same dependence of radiative efficiency on the Eddington ratio, 
then higher-luminosity AGNs will have higher SMBH masses. {\it If} there is a correlation between 
SMBH mass and dark matter halo mass, then the higher-luminosity AGNs will also be more strongly 
clustered.
The observed correlation between the mass of the SMBH and the stellar velocity dispersion in the 
bulge of the galaxy (\citealt{gebhardt_bender_2000}; \citealt{ferrarese_merritt_2000}) suggests 
that a luminosity (X-ray and optical) clustering dependence could be feasible and physically 
motivated, since massive bulges preferentially reside in massive DMHs.
Alternatively, even among galaxies with the same SMBH mass, those that reside in a denser 
large-scale environment may have a higher chance of large mass accretion through galaxy mergers and 
interactions.
However, this trend has not been detected, possibly due in part to the relatively large 
uncertainties in the clustering measurements of X-ray and optically selected AGN samples.  

Several studies have attempted to measure the dependence of clustering on X-ray luminosity.
\cite{gilli_zamorani_2009} compute the clustering signal 
for a high and low X-ray luminosity AGN sample in the {\em XMM-Newton}--COSMOS field at a dividing line 
of $L_{\rm X}^{0.5-10\,{\rm keV}} = 10^{44}$ erg\,s$^{-1}$ and find no significant difference. 
\cite{yang_mushotzky_2006} studied an X-ray luminosity dependence in the {\em Chandra} 
selected AGN samples. As noted by both \cite{yang_mushotzky_2006} and \cite{gilli_zamorani_2009}, 
splitting the samples in different X-ray luminosities leads to the study of a high and low redshift 
sample of X-ray selected AGNs. These studies of the dependence of clustering on luminosity 
are therefore hampered by the redshift-luminosity degeneracy. 
Taking into account the possible redshift evolution of the 
clustering signal, \cite{yang_mushotzky_2006} do not find a significant dependence of
clustering on the X-ray luminosity.  \cite{coil_georgakakis_2009} measure the clustering of AGNs with 
 $10^{42}$ erg\,s$^{-1} < L_{\rm X}^{2-10\,{\rm keV}} < 10^{43}$ erg\,s$^{-1}$  and
 $L_{\rm X}^{2-10\,{\rm keV}} < 10^{42}$ erg\,s$^{-1}$ at similar redshift ($z=0.8$) and find no 
significant difference in their clustering properties as well.

Our sample here covers a limited redshift range, $0.16 <z<0.36$, such that 
we do not expect a significant contribution of the redshift evolution to the clustering 
signal within our sample. In addition, our use of the 
CCF with a large tracer set of LRGs yields an AGN ACF with high precision.   
Splitting the sample into a low and high X-ray luminosity sample at the commonly 
used AGN/QSO dividing line of \mbox{$L_{\rm X}^{0.5-10\,{\rm keV}}= 10^{44}$ erg\,s$^{-1}$} allows 
us to detect an X-ray luminosity dependence of the clustering at the $\sim$2.5$\sigma$ 
level for the first time, in that X-ray luminous RASS-AGNs at low-$z$ cluster more strongly than 
their low luminosity counterparts. 

Although, as mentioned above, the tracer LRG sample is approximately 
volume-limited, in order to clearly test whether the difference in the clustering signal 
between the high and the low $L_{\rm X}$ RASS-AGN sample could be caused by a change of the 
LRG ACF with redshift (which could mimic a luminosity dependence of the inferred RASS-AGN ACF), 
the redshift dependence of
the LRG sample ACF must be analyzed. \cite{zehavi_eisenstein_2005} study dependence of the 
LRG clustering amplitude with luminosity and redshift and split their $0.16<z<0.36$ LRG sample
into $0.16<z<0.23$ and $0.23<z<0.36$ subsamples; they 
find no redshift evolution in the LRG clustering signal. We repeat this test and 
split our LRG sample into two subsamples containing object in redshift bins of $0.16<z<0.26$ and 
$0.26<z<0.36$. We verify that the LRG ACF of both subsamples agree to much better than 1$\sigma$ 
and therefore exclude redshift evolution of the LRG clustering signal as a possible source of the 
detected luminosity dependence of the RASS-AGN clustering.

Optical QSO/AGN samples based on SDSS, 2QZ, and 2SLAQ samples 
yield comparable low uncertainties in the measured ACF 
due to the large sample sizes (up to a few 10,000 objects), allowing 
measurements of the luminosity dependence of clustering in optically selected 
QSO/AGN samples. \cite{mountrichas_sawangwit_2009} compute the CCF at $z<1$ between QSOs and LRGs 
from 2SLAQ, 2QZ, and SDSS to break the redshift-luminosity degeneracy. They find 
little QSO-LRG cross-clustering dependence on QSO luminosity, implying dark matter halo masses 
of $M_{\rm DMH} \sim 10^{13}$ $h^{-1}$ $M_\odot$ approximately independent of QSO luminosity.
Da \^Angela et al.~(2008) confirm that QSOs of different luminosities 
reside in dark matter halos of similar mass ($M_{\rm DMH} \sim 3\times 10^{12}$ $h^{-1}$ $M_\odot$) 
for $z\sim 1.5$. 

Using the optical luminosity in Table~\ref{table:study} and the $\alpha_{\rm ox}$ 
relation between the optical rest-frame flux density at 2500 \AA\ and the X-ray luminosity flux 
density 
at 2 keV measured for 6224 AGNs with broad emission lines from \cite{anderson_margon_2007}, 
we derive the 
median X-ray luminosities of the optical AGN samples. This allows us to compare these results 
with X-ray selected AGN clustering measurements. The optically selected AGN samples have 
median X-ray luminosities of $L_{\rm X}^{0.1-2.4\,{\rm keV}} \sim 3-9 \times 10^{44}$ erg\,s$^{-1}$,
which is very similar to the X-ray luminosity of our high $L_{\rm X}$ RASS-AGN sample.
These samples are shown in Figure~\ref{xagn_b} as gray filled ellipses. 
Contrary to our findings for the X-ray luminosity 
dependence at low-$z$, at higher redshift high luminosity optically selected 
AGNs appear to be {\it less} clustered than 
low luminosity X-ray selected AGNs. This implies that 
high luminosity AGNs at these redshifts lie in less massive dark matter halos than
low luminosity AGNs.
However, {\em Chandra} and {\em XMM-Newton} X-ray selected AGNs (which define the X-ray samples
at higher redshift) contain a significant 
fraction of type II AGNs, while optically selected AGN samples contain mainly type I AGNs with 
broad optical emission lines. Therefore, X-ray and optically selected samples at 
high redshifts probe different kinds of AGNs. Optically selected, high luminosity 
AGN samples at high redshifts and  high $L_{\rm X}$ RASS-AGN samples at low redshifts likely 
contain the same type of AGNs because of both {\em ROSAT}'s soft energy selection being biased 
toward X-ray unabsorbed type I AGNs and the similar X-ray luminosities measured for both samples.
In terms of their clustering properties, 
optically selected AGN samples at higher redshift reside in 
host dark matter halo masses between $M_{\rm DMH} \sim 10^{12}$ $h^{-1}$ $M_\odot$ and
 $M_{\rm DMH} \sim 10^{13}$ $h^{-1}$ $M_\odot$, similar to the full range observed for the low redshift
RASS-AGN samples. Stated differently, the dark matter halo masses of both the low $L_{\rm X}$ and 
high $L_{\rm X}$ RASS-AGN samples are consistent with those of the optically selected, high luminosity 
AGNs at higher redshift (which have relatively large clustering uncertainties).

To summarize the clustering trends seen, 
at low redshift within the RASS-AGN broad-line sample, higher luminosity AGNs cluster more 
strongly than 
their lower luminosity counterparts, while at higher redshifts ($z \gtrsim 0.7$) the 
higher luminosity optically selected AGNs are less clustered than the lower luminosity X-ray
selected AGNs. However, the higher redshift X-ray AGN samples contain a large fraction of
narrow-line AGNs, and so these trends are not necessarily comparing the same kinds of AGNs
at all redshifts and luminosities.  For example, the low $L_{\rm X}$ AGNs studied at high 
redshifts have lower luminosities than the low redshift, low $L_{\rm X}$ RASS-AGNs (see 
Table~\ref{table:study}).
We propose two possible explanations for the AGN clustering trends observed, assuming 
the validity of the $\alpha_{\rm ox}$ connection, i.e., that over a wide range of X-ray 
and optical luminosity, both luminosities are closely connected independent of redshift:

1) The dominant accretion processes that are triggering AGN activity could be 
different at different redshifts and/or halo masses. For example, one possible scenario
is that at high redshift AGNs may be triggered by interactions or mergers between 
galaxies in group environments, while at low redshifts 
internal processes might be responsible. Consequently, due to different 
underlying physics triggering AGN activity, the observed AGN clustering properties 
at low and high redshifts could differ.

2) Alternatively, the underlying physics may not evolve with time, but the 
kinds of AGNs being compared at different redshifts and luminosities
are fundamentally different.  It may be that high luminosity broad-line AGNs generally reside
in lower mass dark matter halos (at least out to $z\sim1$), and within this population 
the brighter AGNs are more clustered. On the other hand, the lower luminosity AGNs probed
in the deep $z\sim1$ surveys, which contain a mix of broad-line and narrow-line
AGNs, may reside in more massive halos. In this scenario, different kinds of AGNs can have 
different triggering mechanisms. The change in the luminosity dependence 
of the clustering signal seen with cosmic time is simply explained by observing different 
kinds of AGNs in the low and high redshift universe with current data sets.

The present clustering measurements do not allow us to accept or reject these hypotheses; 
smaller uncertainties on the observed clustering and detailed model predictions are needed. 
However, it is clear from the above results and discussion that there is not a simple relation
between luminosity and clustering amplitude for AGNs.

\subsection{Comparison to Galaxy Clustering}

Large-scale structure studies have established the dependence of galaxy clustering on 
morphological type, luminosity, color, spectral type, and stellar mass 
(e.g., \citealt{norberg_baugh_2002}; \citealt{madgwick_hawkins_2003}; \citealt{zehavi_zheng_2005}; 
\citealt{meneux_fevre_2006, meneux_guzzo_2009}; \citealt{coil_newman_2008}). 
These quantities are strongly correlated with each other and with clustering amplitude. 
For a comparison with our AGN clustering measurements, we focus on 
the color-dependence of galaxy clustering. 
The last section of Table~\ref{table:study} lists a representative sample of the color-dependent 
clustering measurements of galaxies to $z=1$, where we have listed results for galaxies 
with $L\sim L^{\ast}$, where we use \cite{blanton_hogg_2003}; \cite{norberg_baugh_2002}, and 
\cite{willmer_faber_2006} to determine $L^{\ast}$ for each sample.
The \cite{meneux_fevre_2006} results in Table~\ref{table:study} correspond to galaxies 1 mag
below $M^{\ast}$ but contain more objects and therefore yield more reliable clustering measurements.   

Blue galaxies show a weaker clustering signal than red galaxies over the wide redshift range 
studied here ($0.07 < z < 1.0$).
At low redshifts, results from larger surveys agree with each other and indicate that 
red galaxies are hosted by dark matter halos of slightly higher mass than 
$M_{\rm DMH} \sim 10^{13}$ $h^{-1}$ $M_\odot$, while blue galaxies reside in 
halos of $M_{\rm DMH} \sim 10^{12}$ $h^{-1}$ $M_\odot$. At $z>0.5$ different studies find 
that red galaxies are more strongly  
clustered than blue galaxies, but the published clustering scale lengths are discrepant. 
In terms of inferred dark matter halo masses, our measurements are in agreement with 
\cite{coil_hennawi_2007}, while the results by \cite{meneux_fevre_2006} give dark matter 
halo masses which are too low when evolve into low-$z$ $L^{\ast}$ galaxies.
The errors on the measurements of optical AGN clustering are too large to constrain the
host galaxy type using the clustering measurements; the results are consistent with optical
AGNs being hosted by either blue or red galaxies.  However, \cite{coil_hennawi_2007} find that at 
the 2$\sigma$ level QSOs are clustered like blue galaxies, not red, at $z=1$.

Our clustering measurements strongly exclude the possibility that 
RASS-AGNs reside in the very massive dark matter halos that host LRGs (Figure~\ref{gagn_b}).
Interestingly, our sample of $\sim$1000 low $L_{\rm X}$ RASS-AGNs clusters 
similarly to blue galaxies, while the high $L_{\rm X}$ RASS-AGNs ($n=562$) have the same clustering 
amplitude as red galaxies at similar redshifts. This suggests that RASS-AGNs are
hosted in a mix of higher fraction of blue galaxies and lower fraction of red galaxies.


\section{Conclusions}
Using cross-correlation measurements between RASS-AGNs and 
SDSS LRGs, we measure the projected cross-correlation $w_p(r_p)$,
and from this we derive the real-space auto-correlation function $\xi(r)$ of X-ray selected AGNs 
with broad optical emission lines at $\langle$$z$$\rangle$$=0.25$ with an unprecedented precision. 
Using the  RASS/SDSS cross-identification sample by \cite{anderson_margon_2007}
results in the largest sample of X-ray selected AGNs ($n=1552$) used for 
clustering studies, covering an area of 5468 deg$^2$. We detect a clustering signal at the 
$\sim$11$\sigma$ level and find a correlation length of $r_0=4.28^{+0.44}_{-0.54}$ $h^{-1}$ Mpc 
and $\gamma =1.67^{+0.13}_{-0.12}$, fitting on scales of $r_p=0.3-15$ $h^{-1}$ Mpc. 

We investigate the luminosity dependence of clustering using 
low and high X-ray luminosity subsamples defined using the common 
AGN/QSO dividing line of $L_{\rm X}^{0.5-10\,{\rm keV}} = 10^{44}$\,erg\,s$^{-1}$. 
We detect an X-ray luminosity dependence of the clustering signal at the 
 $\sim$2.5$\sigma$ level, with the brighter sample being more clustered. 
This is contrasted with  clustering measurements of low luminosity X-ray AGNs and 
optically selected QSOs at redshift $z>0.5$,
which suggest that low X-ray luminosity AGNs are more strongly clustered and reside in more massive 
dark matter halos than their high X-ray luminosity counterparts. 
Possible explanations are that (1) different mechanisms trigger AGN activities at different 
redshifts and/or halo masses or (2) at all redshift, low luminosity AGNs and the brightest 
QSOs reside in red galaxies, while most intermediate luminosity AGNs/QSOs are hosted in blue galaxies.

Our low $L_{\rm X}$ RASS-AGN sample exhibits a similar clustering amplitude as 
blue, star-forming galaxies at similar redshifts, 
while the high $L_{\rm X}$ RASS-AGN sample clusters like red galaxies. The total 
RASS-AGN sample is likely dominated by blue host galaxies but includes a fair fraction of 
red host galaxies. 

We show that the auto-correlation function derived from cross-correlation measurements of
$\sim$1500 AGNs and $\sim$50000 LRGs in an area of $\sim$6000 deg$^2$ constrains the auto-correlation 
function at the 10\% level. Although the RASS is the most sensitive X-ray survey 
ever performed, only the most luminous and X-ray unabsorbed (soft) AGNs were detected. The upcoming 
all-sky {\em eROSITA} mission (\citealt{predehl_andritschke_2007}) with its {\em XMM-Newton}-like 
soft and hard energy high sensitivity should detect $\sim$200,000 AGNs with much lower 
X-ray luminosities. The data generated by the mission will allow one to compute clustering measurements 
with a significantly higher accuracy.

To draw meaningful conclusions from AGN clustering measurements using AGN samples selected
at different wavelengths and to address 
the evolution and luminosity dependence of AGN clustering, lower 
uncertainties on the detected clustering signal are required. We describe and successfully show 
in this paper that the cross-correlation function can be used to not only measure the relative
clustering of AGN/galaxy samples but also to determine the auto-correlation 
function with relatively low uncertainties.


\acknowledgments
We thank Richard Rothschild, Alex Markowitz, and Slawomir Suchy
for helpful discussions. We also like to thank Idit Zehavi for providing the values for 
the red/blue SDSS galaxy clustering measurements, as well as Ryan C. Hickox for 
making the median $M_B$ magnitudes for his samples of red and blue galaxies available.
Last but not least we thank the anonymous referee for the very detailed and helpful 
report.

This work has been supported by NASA grant NNX07AT02G, 
CONACyT Grant Cient\'ifica B\'asica No. 83564, and UNAM-DGAPA Grant 
PAPIIT IN110209.

The {\em ROSAT} Project was supported by the Bundesministerium f{\"u}r Bildung 
und Forschung (BMBF/DLR) and the Max-Planck-Gesellschaft (MPG).
Funding for the Sloan Digital Sky Survey (SDSS) has been 
provided by the Alfred P. Sloan Foundation, the Participating 
Institutions, the National Aeronautics and Space Administration, 
the National Science Foundation, the U.S. Department of Energy, 
the Japanese Monbukagakusho, and the Max Planck Society. 
The SDSS Web site is http://www.sdss.org/.

The SDSS is managed by the Astrophysical Research Consortium (ARC) 
for the Participating Institutions. The Participating Institutions 
are The University of Chicago, Fermilab, the Institute for Advanced 
Study, the Japan Participation Group, The Johns Hopkins University, 
Los Alamos National Laboratory, the Max-Planck-Institute for 
Astronomy (MPIA), the Max-Planck-Institute for Astrophysics (MPA), 
New Mexico State University, University of Pittsburgh, Princeton 
University, the United States Naval Observatory, and the 
University of Washington.



\end{document}